\documentclass[aps,prl, tightenlines, floatfix, twocolumn, amsmath, superscriptaddress, showpacs, showkeys]{revtex4-2}
\usepackage{graphicx}
\usepackage{dcolumn}
\usepackage{bm}
\usepackage{float}
\usepackage{hyperref}
\usepackage{bigints}
\usepackage{xcolor}

\begin{document}

\title{Lorentz-Violating Scenarios for the Highest-Energy Photons from GRB 221009A}

\author{Giorgio Galanti}
\email{gam.galanti@gmail.com}
\affiliation{INAF, Istituto di Astrofisica Spaziale e Fisica Cosmica di Milano, Via Alfonso Corti 12, I -- 20133 Milano, Italy}
\affiliation{INFN, Sezione di Pavia, Via Agostino Bassi 6, I -- 27100 Pavia, Italy}

\author{Marco Roncadelli}
\email{marcoroncadelli@gmail.com}
\affiliation{INFN, Sezione di Pavia, Via Agostino Bassi 6, I -- 27100 Pavia, Italy}
\affiliation{INAF, Osservatorio Astronomico di Brera, Via Emilio Bianchi 46, I -- 23807 Merate, Italy}

\date{\today}
\begin{abstract}
A photon at ${\cal E} \simeq 251 \, \rm TeV$ from GRB 221009A was detected by the Carpet collaboration in 2022 using a partial data set. Very recently, Carpet has completed its full data analysis reporting further support for its previous photon now at ${\cal E} = 300^{+ 43}_{- 38} \, {\rm TeV}$. Within standard propagation models, this observation is in strong tension with conventional expectations since such a photon is absorbed by the CMB. Further, we show that this detection is strongly disfavored within the explored scenarios involving axion-like particles (ALPs) alone. Instead, we find that the considered photon is compatible with specific Lorentz invariant violation (LIV) frameworks with the LIV scale obeying in the linear case ${\cal E}_{{\rm LIV}, 1} < 1.22_{-0.22}^{+0.19} \times 10^{21} \, {\rm GeV}$ at $95 \%$ CL and in the quadratic case ${\cal E}_{{\rm LIV}, 2} < 2.03_{-0.22}^{+0.17} \times 10^{13} \, {\rm GeV}$ at $95 \%$ CL. Finally, we outline scenarios where standard photon-ALP oscillations are combined with LIV-induced modifications of photon propagation, which provide a consistent interpretation of the observations of GRB 221009A including the highest energy photons detected by the LHAASO and Carpet collaborations. 
\end{abstract}
 
\maketitle


{\it Introduction} -- The exceptionally bright gamma-ray burst GRB 221009A -- named the brightest of all times (BOAT)~\cite{boat} -- has been detected on October 9, 2022 by the Swift observatory~\cite{swift} and by the Fermi Gamma-ray Burst Monitor (Fermi-GBM)~\cite{fermi}. It lies at redshift $z = 0.151$~\cite{red1,red2,red3}. Concerning its highest energy photons, the observational situation is as follows. More than 5,000 photons have been recorded by the Water Cherenkov Detector Array (WCDA) of the LHAASO collaboration at ${\cal E} > 500 \, {\rm GeV}$ during the first 2,000 s after the Fermi-GBM trigger time (henceforth, the {\it trigger})~\cite{lhaaso1}. Moreover, 142 photon-like events have been registered by the KM2A detector -- also of the LHAASO collaboration -- in the e\-ner\-gy range $3 \, {\rm TeV} \leq {\cal E} \leq 20 \, {\rm TeV}$ over the time lapse  $230 \, {\rm s} \leq t \leq 900 \, {\rm s}$ after the trigger, 8 of which with ${\cal E} > 10 \, {\rm TeV}$~\cite{lhaaso2}. In addition, preliminary evidence for a single photon-like event of energy ${\cal E} \simeq 251 \, {\rm TeV}$ at $t = 4536 \, {\rm s}$ after the trigger has been reported by the Carpet  collaboration~\cite{carpet2}. The Carpet observatory consists of photon detectors, an inner small-area muon (ISAM) detector and four outer large-area muon (OLAM) detectors -- all of which were operative at the time of GRB 221009A -- but the first reported result was based only on the data collected by the ISAM detector over 4536 s~\cite{carpet2,muons}.  

Very recently, the Carpet collaboration has completed the ana\-ly\-sis of the data collected by the whole detector over one day instead of $4536 \, {\rm s}$. The updated result is a single photon-like event of energy ${\cal E} = 300^{+ 43}_{- 38} \, {\rm TeV}$ coincident -- with  chance probability of about $9 \times 10^{- 3}$ -- with GRB 221009A in its arrival direction and time. The probability that this event is a misidentified hadron is about $3 \times 10^{- 4}$~\cite{carpet3}. Furthermore, the Carpet collaboration has shown that the same {\it emitted} power law photon flux which ultimately fits the photons observed by both the WCDA and KM2A detectors is in order-of-magnitude agreement with the Carpet event when extrapolated to much higher energies~\cite{carpet3}. Finally, why has the Carpet event not been observed by LHAASO and HAWC? According to the Carpet collaboration~\cite{carpet3} its photon-like event was close to the limit of the field of view of the LHAASO experiment, whereas the line of sight to GRB 221009A of the HAWC detector was beneath the horizon~\cite{hawc}. Below, we take this result at face value. 

We stress that the observation of photons of ${\cal E} > 10 \, {\rm TeV}$ from GRB 221009A at redshift $z = 0.151$ is in tension with conventional physics for standard emission models.
In fact, they tend to be fully absorbed through the $\gamma \gamma \to e^+ e^-$ process by the {\it extragalactic background light} (EBL) -- namely the infrared/optical/ultraviolet photons emitted by all galaxies during their whole evolution -- as well as by the CMB and the radio background~\cite{breitwheeler,heitler,nishikov1962,gouldschreder1967,stecker1969,faziostecker1970,stecker1992,madaupozzetti2000,hauwserdwek2001,primack2001,kneiske2002,kneiske2004,primack2005,schroedter2005,aharonian2006,mazinraue2007,mazingoebel2007,frv2008,finkerazzaque2009,gilmore2009,finke2010,kneiskedole2010,dominguez2011,orrkrennrichdwek2011,gilmore2012,dwekkrennrich,inoue2013,dgr2013,stecker2016,franceschinirodighiero2017,ciber2017,gprt2020,saldanalopez2021}, with the CMB becoming the dominant absorption process at the Carpet energies~\cite{stecker1969,faziostecker1970}.  As a consequence, the observation of the highest energy photons from GRB 221009A challenges standard propagation models -- as also stressed in~\cite{lhaaso2} and~\cite{aronkus} for LHAASO photons -- and provides clues of new physics. Such a situation is further exacerbated for the Carpet event.

Elsewhere, we have shown that the 8 photon-like events of ${\cal E} > 10 \, {\rm TeV}$ detected by LHAASO yield a hint at the existence of a viable axion-like particle 
(ALP). 
In addition, we have demonstrated that such a result {\it cannot} be obtained by Lorentz invariance violation (LIV)~\cite{noi}.

Here, relying upon the newly reported Carpet result, we demonstrate that for the ${\cal E} = 300^{+ 43}_{- 38} \, {\rm TeV}$ photon-like event its observability is difficult to explain with ALPs but {\it can} with LIV in specific settings. We note that so far only lower bounds on the LIV scale have been derived by employing -- among other methods -- the very-high-energy (VHE, $100 \, {\rm GeV} < {\cal E} < 100 \, {\rm TeV}$) photon emission from astronomical sources in general, and since 1998 from {\it gamma-ray bursts} (GRBs) in  particular~\cite{limitiLIV1,limitiLIV2,steckerglashowLIV,limitiLIV3,limitiLIV4,limitiLIV5,limitiLIV6,limitiLIV7,limitiLIV8,limitiLIV8a,limitiLIV9,limitiLIV9a,limitiLIV10,limitiLIV11,limitiLIV12,LIVlimLang,limitiLIV13,limitiLIV14,limitiLIV15,limitiLIV16,limitiLIV17}. 

The aim of this Letter is to show under what assumptions specific LIV models {\it do} explain the Carpet event. More generally, we offer scenarios in which photon-ALP oscillations occur in certain LIV frameworks so that a consistent interpretation of the observations of GRB 221009A emerges including the highest energy photons detected by the LHAASO collaboration. We emphasize that much more details and all unproved statements (for lack of space) made in this Letter are justified in the Supplemental Material~\cite{SupMat}.

\

{\it General strategy} -- In order to carry out a quantitative analysis of the Carpet event, we need to know the emitted photon flux ${\cal F}_{\rm em}({\cal E})$. As already stated, it has been shown by the Carpet collaboration that a good fit to ${\cal F}_{\rm em}({\cal E})$ is given by a higher-energy extrapolation of the one reported by LHAASO~\cite{lhaaso1,lhaaso2} (see Fig. 7 of~\cite{carpet3}). The expected photon flux ${\cal F}_{\rm exp}({\cal E})$ is obtained upon multiplication of ${\cal F}_{\rm em}({\cal E})$ by the photon survival probability $P({\cal E}; \gamma \to \gamma)$ which quantifies the EBL, CMB and radio absorption, as well as new physical effects~\cite{noz}. Yet, by definition the expected photon flux is ${\cal F}_{\rm exp} ({\cal E}) \equiv d N_{\gamma} ({\cal E})/(d {\cal E} \, d A \, dt)$ where $d N_{\gamma} ({\cal E})$ is the number of expected photons in the energy range $d {\cal E}$ about 
${\cal E}$ within the effective detector area $d A$ during the exposure time $d t$. Hence we get
\begin{equation}
\frac{d N_{\gamma} ({\cal E})}{d {\cal E} \, d A \, d t} = P({\cal E}; \gamma \to \gamma) {\cal F}_{\rm em}({\cal E})~. 
\label{16112025a}
\end{equation}
So -- when $P({\cal E}; \gamma \to \gamma)$ is known -- we obtain the number of expected  Carpet photons $N_{\gamma}$ by integrating Eq. (\ref{16112025a}) over the Carpet energy range $262 \, {\rm TeV} \leq {\cal E} \leq 343 \, {\rm TeV}$ and multiplying it by both the Carpet effective area of $\sim 60 \, \rm  m^2$ and the exposure time of one day~\cite{carpet3}.

{\it Conventional physics} -- $P_{\rm CP} ({\cal E}; \gamma \to \gamma)$ is computed in the Supplemental Material~\cite{SupMat}, and the above procedure based on Eq. (\ref{16112025a}) gives $N_{\gamma} ({\rm CP}) \sim 10^{- 96}$. We restate that at this energy the leading photon absorption is due to the CMB, which makes our result  robust since its uncertainty is much smaller than that affecting the EBL. 

{\it Axion-like particles (ALPs)} -- They are very light neutral pseudoscalar bosons (see Supplemental Material~\cite{SupMat}) quite similar to the axion~\cite{axion1,axion2,axion3}, but their mass $m_a$ and two-photon coupling $g_{a \gamma \gamma}$ are {\it unrelated}. ALPs are attracting an ever growing interest since they are a generic prediction of superstring theory (for a detailed motivation see~\cite{ringwald1,ringwald2} and for a very incomplete list of references  see~\cite{turok1996,string1,string2,string3,string4,string5,axiverse,abk2010,cicoli2012,dias2014,cicoli2014a,conlon2014,cicoli2014b,conlon2015,cicoli2017,scott2017,cisterna1,cisterna2}). Moreover, they explain some astrophysical anomalies~\cite{trgb2012,grdb,noi} and are among the best dark matter candidates~\cite{arias} (general reviews on ALP are~\cite{irastorzaredondo,gruniverse,GalantiRev,GalantiRevBlazar}). Because we are interested in their interactions with photons alone, the corresponding Lagrangian is~\cite{natun} 
\begin{equation}
{\cal L}_{a \gamma \gamma} = - \, \frac{1}{4} \, g_{a \gamma \gamma} \, F_{\mu \nu} \, {\tilde{F}}^{\mu \nu} \, a = g_{a \gamma \gamma} \, {\bf E} \cdot {\bf B} \, a~,
\label{a2}
\end{equation}
where $a$ is the ALP field, while ${\bf E}$ and ${\bf B}$ are respectively the electric and magnetic components of the electromagnetic tensor $F_{\mu \nu}$ whose dual is ${\tilde{F}}^{\mu \nu}$. 
ALP-induced astrophysical effects arise in the presence of an {\it external} ${\bf B}$ field, and so ${\bf E}$ is the {\it electric} field of a {\it propagating} photon. Therefore, the last term in Eq. (\ref{a2}) yields an off-diagonal mass matrix in the photon-ALP sector so that the interaction eigenstates differ from the mass eigenstates. This fact gives rise to two effects~\cite{mpz,rs}: 1) photon-ALP oscillations, and 2) change of the photon polarization state, which are potentially detectable in high-energy and VHE astrophysics both on astrophysical spectra~\cite{drm2007,simet2008,sanchezconde2009,dgr2011,trgb2012,wb2012,trg2015,kohri2017,galantironcadelli20118prd,grjhea,gtre2019,grdb,noi} and on photon polarization~\cite{ALPpol1,bassan,ALPpol2,ALPpol3,ALPpol5,day,galantiTheo,galantiPol,grtcClu,grtBlazar}. The key point is that photon-ALP oscillations {\it largely reduce} the EBL, CMB and radio absorption, thereby allowing a {\it considerably greater} number of VHE photons to be expected (see Supplemental Material~\cite{SupMat}). Moreover, when ${\cal E} \gg m_a$ -- which is presently the case -- the photon-ALP beam propagation equation along the $y$-axis becomes~\cite{rs}
\begin{equation}
\label{02012026a} 
\left(i \, \frac{d}{d y} + {\cal E} +  {\cal M} ({\cal E},y) \right)  \left(\begin{array}{c}A_x (y) \\ A_z (y) \\ a (y) \end{array}\right) = 0~.
\end{equation}
Here $A_x (y)$ and $A_z (y)$ denote the two photon polarization amplitudes along the $x$ and $z$ axes, respectively, and $a (y)$ the ALP amplitude. ${\cal M} ({\cal E},y)$ is the photon-ALP mixing matrix (see Supplemental Material~\cite{SupMat}) which fails to be self-adjoint when photon absorption occurs. Since Eq. (\ref{02012026a}) is a Schr\"odinger-like equation (with $t \to y$) it follows that the beam is formally described as a {\it non-relativistic 3-level decaying quantum system}, a fact that greatly simplifies the calculations. 

Now, we are in a position to investigate whether photon-ALP oscillations alone can explain the Carpet event. We evaluate $P_{\rm ALP} ({\cal E}; \gamma \to \gamma)$ by accounting for the relevant astrophysical uncertainties (see Appendix and Supplemental Material~\cite{SupMat}). To maintain our previous explanation of the LHAASO photons~\cite{noi}, the ALP parameters are constrained to $m_a \simeq (10^{- 11} - 10^{- 7}) \, {\rm eV}$ and $g_{a \gamma \gamma} \simeq (3 - 5) \times 10^{- 12} \, {\rm GeV}^{- 1}$. Across this entire parameter space, the expected photon count for the Carpet event remains $N_{\gamma} ({\rm ALP}) \lesssim {\cal O}(10^{-4})$. Since $N_{\gamma} ({\rm ALP})$ is manifestly a very small number in the Carpet case we have to use the Poisson statistics. Accordingly, at $95 \%$ CL we should satisfy $N_{\gamma} ({\rm ALP}) \geq 0.0513$~\cite{gehrels}, and so ALPs fail to bridge the gap to the observation by about two orders of magnitude. 

\  

{\it Lorentz invariance violation (LIV)} -- Manifestly, a new kind of physics has to be looked for in order to allow for the observability of the Carpet event, which must coexist with the previous ALPs in such a way to retain their explanation of the LHAASO VHE photons. To this end, we suppose that LIV is a good option. We first investigate whether LIV alone accomplishes the desired task, and next we address the merging of ALPs and LIV into consistent scenarios. Even though LIV theory is well known~\cite{amelino1,libmac2009,mattingly,amelino2,liberati,gtl,addazi,batista}, for the reader's convenience we very briefly recall its implications that are relevant for us. The widespread belief is that gravity should be quantized, but this goal has not yet been achieved. Nonetheless, its basic features have clearly emerged. Quantum gravitational effects show up around the Planck scale $M_P \equiv (\hbar c/G)^{1/2} \simeq 1.22 \times 10^{19} \, {\rm GeV}$ where they give rise to an ever changing metric and topology which impart a {\it foam-like structure} to spacetime~\cite{wheeler1,wheeler2,wheeler3,hawking1978,carlip1997}. A natural expectation is that the resulting particle propagation exhibits a violation of Lorentz invariance at a scale ${\cal E}_{\rm LIV}$ -- which can be smaller or larger than $M_P$ by some orders of magnitudes -- so that the goal of LIV theories is to capture its low-energy relic effects in special regimes. One of them is the deformation of the photon dispersion relation $p^2 = {\cal E}^2$. A convenient parametrization is 
\begin{equation}
p^2 = {\cal E}^2 \, \left[1 + f \left(\frac{{\cal E}}{{\cal E}_{\rm LIV}} \right) \right]~,
\label{liv}
\end{equation}
where $f ( \cdot )$ is a model-dependent smooth function such that $f (0) = 0$ since it must vanish in the limit ${\cal E}_{\rm LIV} \to \infty$. At energies ${\cal E} \ll {\cal E}_{\rm LIV}$ Eq. (\ref{liv}) can be expanded in a power series of ${\cal E}/{\cal E}_{\rm LIV}$. A drawback of this approach is that some terms must be dropped when they give rise to unwanted physical effects, since the coefficients of the expansion cannot be predicted. Denoting as ${\cal E}^n/{\cal E}_{{\rm LIV}, n}^n$ the first physically allowed term, Eq. (\ref{liv}) reduces to 
\begin{equation}
p^2 = {\cal E}^2 \, \left(1 + \xi_n \, \frac{{\cal E}^n}{{\cal E}_{{\rm LIV}, n}^n} \right) 
\label{liv2}
\end{equation}
with $\xi_n = 1$ corresponding to subluminal propagation and $\xi_n = -1$ to superluminal propagation while ${\cal E}_{{\rm LIV}, n}$ denotes the LIV energy scale at order $n$. Observe that formally the LIV term looks like an effective photon mass square $m^2_{\rm eff} \equiv - \, \xi_n \, {\cal E}^{n + 2}/{\cal E}^n_{{\rm LIV}, n}$. In the following, we shall restrict our attention to the {\it subluminal case} and to the two leading orders, $n=1, 2$. LIV modifies the quantum mechanical propagators and the reaction thresholds. Our focus is on the implications of these effects for the $\gamma \gamma \to e^+ e^-$ process~\cite{gonzales,coleman,kifune1999,aloisio2000,amelinopiran}. At Carpet energies it turns out that for subluminal propagation VHE photons from a cosmological source interact with background photons at {\it higher} energies where the photon density is {\it lower} thereby increasing the transparency of the Universe, which is quantified by the resulting photon survival probability~\cite{steckerglashow,stecker2003,jacobpiran}. Thus, it looks natural to inquire whether such an effect allows the Carpet event to be expected (for superluminal propagation the cosmic transparency is lower than in conventional physics, and so we discard this possibility). Recall that at ${\cal E} \sim 300 \, \rm TeV$ the standard cosmic transparency is essentially determined by the CMB~\cite{stecker1969,faziostecker1970}, hence making our subsequent LIV bounds remarkably robust against the uncertainties affecting EBL models, which instead dominate at lower energies. Another LIV effect which follows from Eq. (\ref{liv2}) is that photons acquire an energy-dependent velocity~\cite{addazi,batista,limitiLIV1}. Hence, for subluminal propagation they are affected by a {\it time delay}~\cite{timedelay} and are stable against photon decay ($\gamma \to e^+ e^-$) and photon splitting ($\gamma \to N \gamma$)~\cite{addazi,batista}. So far, we have considered LIV effects for photons but LIV can affect any particle. Is the LIV modification of the dispersion relations of photons and massive particles the same? As classical gravity acts equally on any particle (equivalence principle), the answer seems to be yes. However, this is not true in general since LIV-induced quantum gravitational effects depend on the considered framework. For instance, a LIV acting differently on electrons and photons has been found in the Liouville string model of spacetime foam~\cite{ellis1,ellis2,ellis3,ellis4}, where only photons have a LIV-deformed dispersion relation. More generally, a non-universal LIV emerges naturally in the D-brane physics since excitations carrying nonvanishing charges under the Standard Model gauge group are represented as open strings with their ends attached to a brane, while neutral excitations propagate as closed strings in the bulk space transverse to the brane~\cite{brane}: only the latter excitations -- like photons and ALPs -- can possess LIV-deformed dispersion relations.   

\

{\it New consistent scenarios} -- Hitherto we have dealt with LIV alone. Now we have to identify the contexts in which the LIV-induced observation of the Carpet event is consistent with the ALP-induced detection of the ${\cal E} > 10 \, {\rm TeV}$ LHAASO photons. We assume that photon-ALP oscillations are Lorentz invariant also in the presence of LIV. As a consequence, two conditions should be met: 1) the term ${\bf E} \cdot {\bf B} \, a$ in Eq. (\ref{a2}) is {\it not} affected by LIV, and 2) photons and ALPs violate Lorentz invariance by the {\it same} amount. We stress that the emerging frameworks arise from a LIV effective Lagrangian containing only two terms with equal coefficients: one for the photon and the other for the ALP (they are dimension 5 operators in the $n = 1$ case and dimension 6 operators in the $n = 2$ case)~\cite{liberati}. The resulting implications for the beam propagation equation (\ref{02012026a}) are discussed in the Supplemental Material~\cite{SupMat}, along with the calculation of $P_{{\rm ALP + LIV}, n} ({\cal E}; \gamma \to \gamma)$ whose LIV contribution is evaluated as in~\cite{kifune1999,fairbairn2014,tavLIV2016} for the deformed dispersion relation~(\ref{liv2}) in the  cases $n=1, 2$. In the Supplemental Material~\cite{SupMat} it is also shown that consistent ${\rm ALP} + {\rm LIV}$ scenarios emerge. An intriguing possibility is that the interaction term in Eq. (\ref{a2}) arises as an effect of quantum corrections within an underlying LIV background~\cite{petrov1,petrov2}. 

\

{\it Results} -- Our main task is the determination of ${\cal E}_{{\rm LIV}, n}$ for $n=1, 2$  in such a way that one photon-like event of ${\cal E} = 300^{+ 43}_{- 38} \, {\rm TeV}$ is observed from GRB 221009A. We follow the same strategy outlined just after Eq.~(\ref{16112025a}) to evaluate the expected number of photons $N_{\gamma}  ({\cal E}_{{\rm LIV}, n})$ which initiate the single air-shower event observed by Carpet. As in the ALP case, we have to employ the Poisson statistics since $N_{\gamma} ({\cal E}_{{\rm LIV}, n})$ is a very small number. Correspondingly, at $95 \%$ CL we should satisfy $N_{\gamma} ({\cal E}_{{\rm LIV}, n}) \geq 0.0513$~\cite{gehrels}. We can now go back to Eq.~(\ref{16112025a}), which can be rewritten as 
\begin{equation}
\frac{d N_{\gamma} ({\cal E})}{d {\cal E} \, d A \, d t} = 
P_{{\rm ALP + LIV}, n} ({\cal E}; \gamma \to \gamma) {\cal F}_{\rm em}({\cal E})~. 
\label{12022026q}
\end{equation}
We solve this equation in the same fashion as for Eq. (\ref{16112025a}), by integrating over the Carpet energy range and multiplying by the effective area and exposure time. Hence, by imposing the condition $N_{\gamma} ({\cal E}_{{\rm LIV}, n}) \geq 0.0513$ we get the upper bounds on the LIV scales at $95 \%$ CL 
\begin{equation}
{\cal E}_{{\rm LIV}, 1} < 1.22_{-0.22}^{+0.19} \times 10^{21} \, {\rm GeV}~, \ \ \ \ \ \ n=1~, 
\label{12022026w} 
\end{equation}
\begin{equation}
{\cal E}_{{\rm LIV}, 2} < 2.03_{-0.22}^{+0.17} \times 10^{13} \, {\rm GeV}~, \ \ \ \ \ \ n=2~,
\label{12022026w2}
\end{equation}
where the reported errors reflect the systematic uncertainties on the Carpet spectral normalization~\cite{carpet3} (see Appendix for details). 

As carefully shown in the Supplemental Material~\cite{SupMat}, the above limits are in agreement with all available bounds for $n = 1$ and $n = 2$ LIV. Nevertheless, one of them deserves to be mentioned here. In conventional physics $n = 1$ LIV gives rise to birefringence (that is a rotation of the polarization plane of photons) hence astronomical sources at cosmological distances like GRBs should be unpolarized~\cite{laurent2011}. This is in sharp contradiction with astronomical evidence, and the most updated bounds are ${\cal E}_{{\rm LIV}, 1} > 3.6 \times 10^{34} \, {\rm GeV}$ and ${\cal E}_{{\rm LIV}, 2} > 1.3 \times 10^{11} \, {\rm GeV}$~\cite{addazi,covino,vasileiou}. While this looks at odd with our result for $n = 1$, it becomes viable in the above-considered D-brane models since -- in spite of the fact that the photon dispersion relation is generally LIV-deformed -- no birefringence arises because the two circular polarization states of the photon propagate at the same speed~\cite{libmac2009,not}. No similar limitation exists for the $n = 2$ case. 

Moreover, Ofengeim and Piran~\cite{PiranLIV2} have very recently demonstrated that the time delay of more than one hour between the observation of VHE LHAASO photons and the Carpet event is explained by $n = 2$ LIV for ${\cal E}_{{\rm LIV}, 2} = 1.59_{- 0.35}^{+ 0.56} \times 10^{12} \, {\rm GeV}$. Observe that this value of ${\cal E}_{{\rm LIV}, 2}$ is {\it consistent} with our upper bound. 

\begin{figure}
\begin{center}
\includegraphics[height=10.3cm]{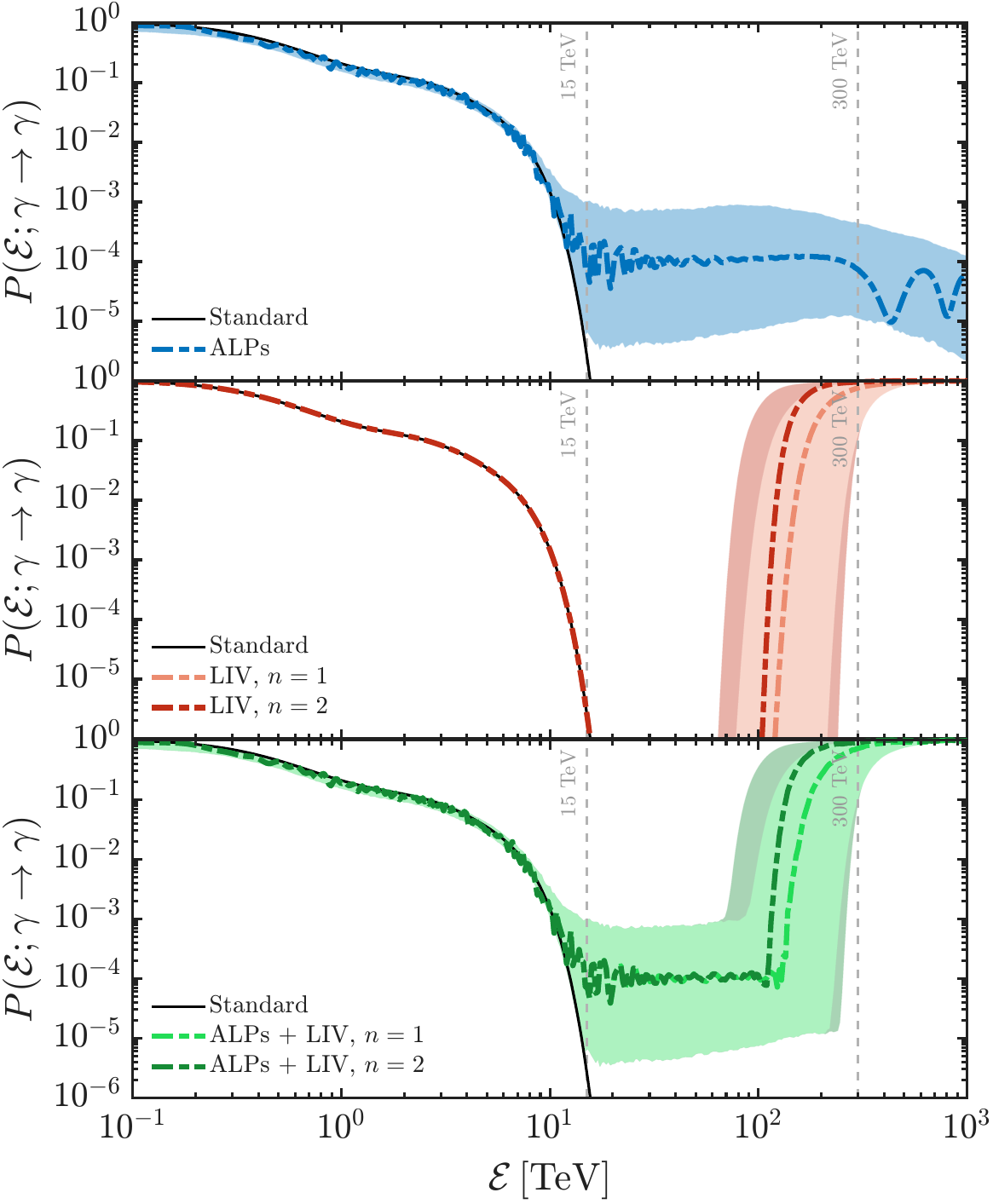}
\end{center}
\caption{\label{probFig} Photon survival probability $P ({\cal E}; \gamma \to \gamma)$ versus energy ${\cal E}$, taking into account ALP and LIV effects separately (upper and central panels) and in combination (lower panel). Curves correspond to our benchmark values, while shaded bands illustrate the range of predicted values for the corresponding scenarios (lighter bands for LIV with $n=1$, darker bands for $n=2$) as obtained from the parameter variations described in the Appendix. Conventional physics is shown in all panels for comparison.}
\end{figure}

\begin{figure}
\begin{center}
\includegraphics[height=10.3cm]{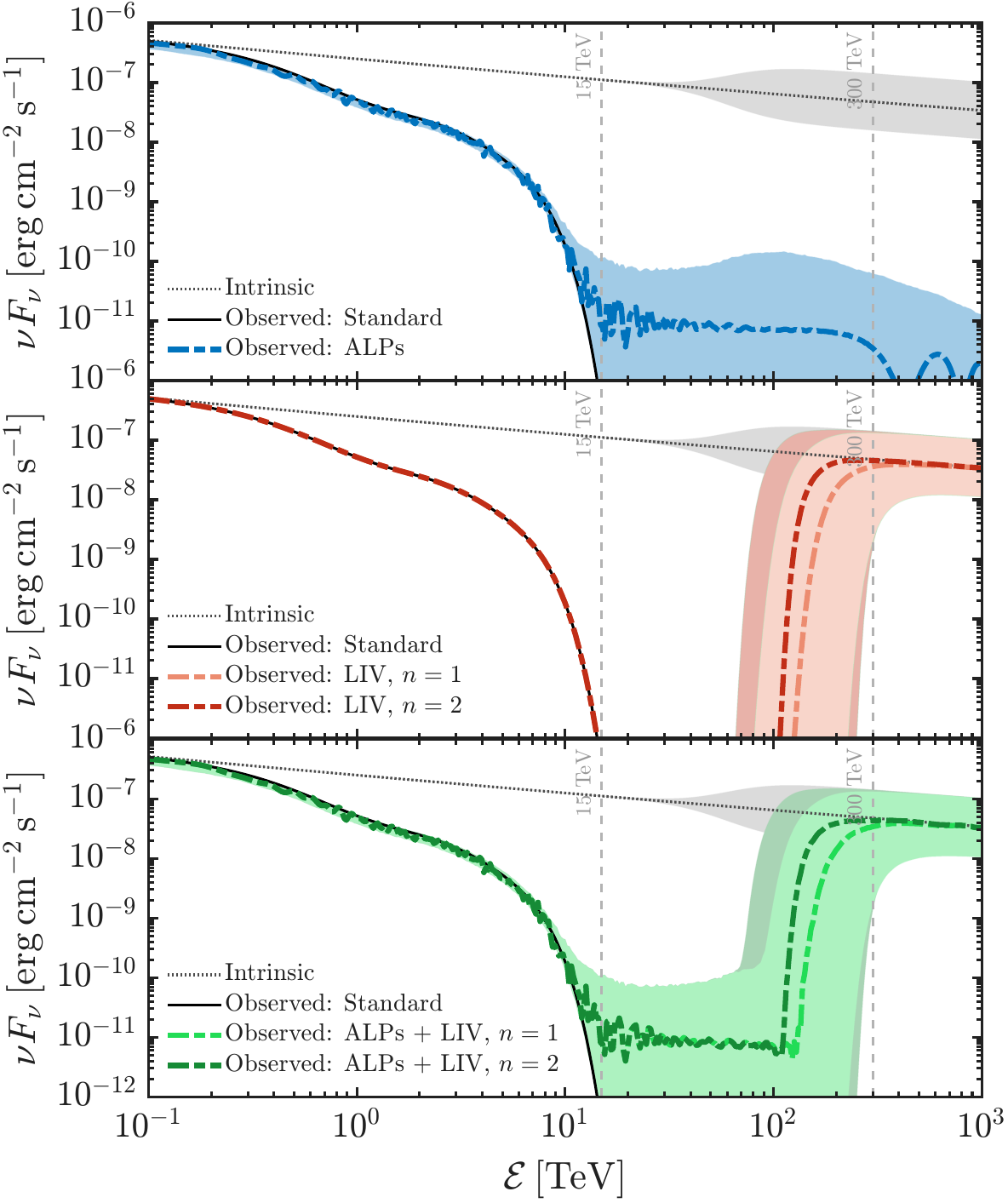}
\end{center}
\caption{\label{spectrumFig} Expected SED versus energy ${\cal E}$, taking into account ALP and LIV effects separately (upper and central panels) and in combination (lower panel). Curves correspond to our benchmark values, whereas shaded bands illustrate the range of predicted SED for the corresponding scenarios (lighter bands for LIV with $n=1$, darker bands for $n=2$), as obtained from the parameter variations described in the Appendix. The black dotted line denotes the intrinsic SED of GRB 221009A as estimated by LHAASO~\cite{lhaaso1,lhaaso2} and extrapolated to Carpet energies~\cite{carpet3} with its associated uncertainty shown as a shaded region. Observed SED versus energy ${\cal E}$ in conventional physics is reported in all panels for comparison.}
\end{figure}

A full account of our results is exhibited in Figs.~\ref{probFig} and~\ref{spectrumFig}, where we plot the photon survival probability and the expected spectral energy distribution [SED $\equiv \nu \, F_{\nu} ({\cal E}) = {\cal E}^2 \, {\cal F}_{\rm exp} ({\cal E})$] in the various  cases. 

As benchmark values, we take $m_a = 10^{-10} \, \rm eV$, $g_{a\gamma\gamma}= 4 \times 10^{-12} \, \rm GeV^{-1}$, ${\cal E}_{{\rm LIV},1} = 3 \times 10^{20} \, {\rm GeV}$, and ${\cal E}_{{\rm LIV},2} = 5 \times 10^{12} \, {\rm GeV}$, all consistent with the above constraints. In Fig.~\ref{spectrumFig} we include the uncertainty on the intrinsic Carpet spectrum. In Figs.~\ref{probFig} and~\ref{spectrumFig}, the colored regions illustrate the range of predicted photon survival probability and, respectively, SED in the ALP-only, LIV-only, and ALP + LIV scenarios, as obtained by varying the relevant parameters within their allowed ranges (see Appendix for details), indicating that the qualitative behavior is stable across the explored configurations.

\

{\it Discussion and Conclusions} -- The GRB 221009A is unique in several respects. Not only is it the brightest GRB of all times -- to such an extent to ionize the upper ionosphere -- but also the one whose emitted photons have reached the highest energies. As shown elsewhere~\cite{noi}, the photons of ${\cal E} > 10 \, {\rm TeV}$ detected by the LHAASO collaboration~\cite{lhaaso2} provide a hint at the existence of an ALP with mass $m_a \simeq (10^{- 11} - 10^{- 7}) \, {\rm eV}$ and two-photon coupling $g_{a \gamma \gamma} \simeq (3 - 5) \times 10^{- 12} \, {\rm GeV}^{- 1}$. In this Letter, we have interpreted the photon-like event of ${\cal E} = 300^{+ 43}_{- 38} \, {\rm TeV}$ observed by the Carpet collaboration~\cite{carpet3} as a LIV effect. 
 
Within conventional physics an additional emission component -- such as proton-synchrotron or of hadronic origin -- from GRB 221009A is insufficient to account for the Carpet result since it could mitigate the problem by a few orders of magnitude, but a correction of more than 90 orders of magnitude would be compelling. We have also shown that ALPs cannot account for the GRB 221009A observation at ${\cal E} = 300^{+ 43}_{- 38} \, {\rm TeV}$ within the considered parameter space. In addition, we find no qualitative change in the main conclusions across the full range of model parameters explored in this analysis when the uncertainties associated with the Carpet spectrum are included.

Besides, it looks tantalizing that the combination of our result with that of Ofengeim and Piran~\cite{PiranLIV2} shows that $n$ = 2 LIV in four-dimensional space accounts for both the Carpet photon-like event and its time delay with respect to the VHE LHAASO ones. 

Before closing this Letter, there is one more point worth mentioning. Our consistent scenarios are not {\it ad hoc}, in that ALPs explain not only the VHE photons detected by LHAASO but also two other astrophysical anomalies~\cite{trgb2012,grdb}. As systematically discussed in the Supplemental Material~\cite{SupMat}, this represents a unique feature of our proposal as compared to others. In addition, the alternative explanations put forward so far face significant challenges. Indeed, unconventional attempts based on standard physics such as the {\it proton beam model}~\cite{aronkus} and the {\it neutron beam model}~\cite{carpet3,neutronbeam} appear unable to account for the observation within a realistic physical setup (see Supplemental Material~\cite{SupMat}). 

In conclusion, the combined ${\rm ALP} + {\rm LIV}$  frameworks presented in this Letter provide a consistent interpretation of the GRB 221009A VHE emission, suggesting a potential first indication of LIV in the considered contexts. Needless to say, our findings require further confirmations by future observations. Only time will tell whether the above two clues are real discoveries.

\

{\it Data Availability Statement} -- The data that support the findings of this study are available from the corresponding author upon request.

\

{\it Acknowledgments} -- We thank the Referees for their constructive suggestions. We also thank Giacomo Bonnoli, Lara Nava and Fabrizio Tavecchio for their comments on a preliminary draft of this Letter, and Andrea Belfiore, Andrea De Luca, Giovanni Pareschi for discussions. The work of G. G. is supported by a contribution from Grant No. ASI-INAF 2023-17-HH.0 and by the INAF Mini Grant `High-energy astrophysics and axion-like particles', PI: G. G, and that of M. R. by an INFN grant.

\

{\it Appendix} -- We describe the procedure used to construct the bands shown in Figs.~\ref{probFig} and~\ref{spectrumFig} for the ALP-alone, LIV-alone, and ALP + LIV scenarios. The three cases rely on different modeling assumptions: the ALP scenario is obtained from a systematic scan over the relevant parameter space, the LIV scenario is constrained by physically allowed intervals bounded by the independent upper bounds from Eqs. (\ref{12022026w}) and (\ref{12022026w2}) and lower limits from observations, and the combined ALP + LIV case is derived from a single consistent propagation model in which both effects are simultaneously included within their respective allowed parameter ranges. 

We emphasize that this $95\%$ inclusion construction is distinct from the $95\%$ confidence level used in the Poisson inference for the LIV bounds.

In all cases, we include the uncertainty associated with the intrinsic Carpet spectrum at ${\cal E}\sim 300 \, \rm TeV$, allowing for a factor-of-three variation above and below the extrapolated LHAASO flux, in line with the systematic uncertainty reported by the Carpet analysis~\cite{carpet3} (see also Fig.~\ref{spectrumFig}). 

\

\noindent {\it Robustness and limits of the ALP scenario}: The range of ALP parameters $m_a$ and $g_{a\gamma\gamma}$ adopted in the main text ensures $P_{\rm ALP}({\cal E};\gamma\to\gamma)\gtrsim \mathcal{O}(10^{-5})$ at ${\cal E}\sim 15 \,\rm TeV$, as required to account for the LHAASO data~\cite{noi}. This choice incorporates the main astrophysical uncertainties, including host-galaxy morphology, magnetic-field configurations, and EBL models (see Supplemental Material~\cite{SupMat}). In particular, a higher EBL density or weaker magnetic fields in the host galaxy require values of $g_{a\gamma\gamma}$ toward the upper end of the considered range. Conversely, $m_a$ is constrained to $m_a \lesssim 10^{-7} \,\rm eV$ by the efficiency of photon-ALP conversion and to $m_a \gtrsim 10^{-11} \,\rm eV$ by X-ray bounds~\cite{limJulia}. We have also verified that including photon-ALP conversion in extragalactic space does not significantly affect the results (see Supplemental Material~\cite{SupMat}). 

The colored bands in Figs.~\ref{probFig} and~\ref{spectrumFig} are obtained by scanning the relevant parameter space (including $m_a$, $g_{a\gamma\gamma}$, and magnetic-field configurations), generating $\sim 10^3$ realizations. The bands correspond to the central $95\%$ inclusion region, defined by excluding the lowest and highest $2.5\%$ of the realizations at each energy. 

We find that variations in the EBL model and in the magnetic field of the host galaxy do not qualitatively modify the results. In particular, for ALPs at ${\cal E}\sim 300 \,\rm TeV$, the expected photon count remains around two orders of magnitude below the level required to account for the observation at $95\%$ CL, even when considering the most favorable configurations within the explored parameter space (i.e., maximal coupling, magnetic field strength, and coherence length). This demonstrates the robustness of the conclusion that ALPs alone cannot account for the Carpet event. 

\

\noindent {\it Robustness and constraints of the LIV scenario}: For the LIV case, the bands correspond to the range of LIV scales consistent with both our derived upper limits, obtained from the requirement to explain the Carpet event at $95\%$ CL (including the uncertainty on the observed flux), and the most stringent lower bounds available in the literature relevant to this scenario~\cite{LIVlimLang}. The resulting region therefore reflects the physically allowed interval for ${\cal E}_{{\rm LIV},n}$ propagated to the corresponding observables, rather than a statistical confidence band.

In constructing the bands, we adopt the most restrictive upper limit on ${\cal E}_{{\rm LIV},n}$ in Eqs. (\ref{12022026w}) and (\ref{12022026w2}) including the uncertainty in the Carpet flux. This choice defines the most conservative case compatible with the observation; alternative choices within the allowed range produce only marginal variations in the resulting bands.

Additional LIV constraints are discussed in the Supplemental Material~\cite{SupMat}.

At the energies relevant for the Carpet event (${\cal E}\sim 300 \,\rm TeV$), the dominant attenuation mechanism in conventional physics is provided by interactions with the CMB, while the role of the EBL is subleading~\cite{stecker1969,faziostecker1970}. As a consequence, the determination of the LIV scale is largely insensitive to uncertainties in EBL modeling. Moreover, since LIV effects enter through modifications of the photon dispersion relation and of the pair-production threshold, they do not directly depend on astrophysical parameters such as magnetic-field configurations. This is in contrast with the ALP scenario, where magnetic-field properties play a central role in the conversion probability. 

\

\noindent {\it ALP + LIV scenario}: For the combined ALP + LIV case, the bands are obtained from a fully consistent propagation model in which photon-ALP oscillations and LIV-modified dispersion relations are included simultaneously in the beam evolution. We perform the same scan over the ALP parameter space as described above, while for each realization the LIV scale ${\cal E}_{{\rm LIV},n}$ is varied within the physically allowed range defined by our upper limits (from the requirement of explaining the Carpet event at $95\%$ CL) and by the most stringent lower bounds available in the literature relevant to this case.

The resulting regions therefore represent the envelope of the predictions of the complete ALP + LIV framework, obtained from an independent variation of all parameters within their physically allowed domains. We emphasize that this construction does not correspond to a superposition of the separate ALP and LIV effects, but arises from a single propagation calculation in which both mechanisms are simultaneously active.

\

\begin{widetext}

\vskip 1 cm

\section{SUPPLEMENTAL MATERIAL}

\medskip
\medskip

\section{Properties of the host galaxy}

Observations with both HST (Hubble space telescope) and JWST (James Webb space telescope) have shown that GRB 221009A is located at about 0.65 kpc from the centre of a star-forming spiral galaxy observed edge-on in first approximation~\cite{levan,blanchard}. As we shall see, this fact is of crucial importance for our considerations. Emission lines of 
${\rm H}_2$ -- which trace dense star-forming regions -- have been detected from several patches of the host but the strongest ones from the site of GRB 221009A. The 
star-formation-rate is ${\rm SFR} = 0.17 \, M_{\odot}/{\rm yr}$~\cite{blanchard}, and so the host is likely to be a rather normal spiral. 

\

\section{Maximal LHAASO photon energy}

The maximal photon energy estimated by the LHAASO collaboration depends on the assumed spectral function used to fit the data, resulting in $17.8_{- 5.1}^{+ 7.4} \, {\rm TeV}$ for a log-parabola (LP) and $12.5_{- 2.4}^{+ 3.2} \, {\rm TeV}$ for a power-law with an exponential cutoff (PLEC)~\cite{lhaaso2}. Note that in the latter case the highest energy point has an unphysical turn up, and so the former case seems preferable.
Specifically, the LHAASO collaboration considers two time intervals: $230 \, {\rm s} < t < 300 \, {\rm s}$ and $300 \, {\rm s} < t < 900 \, {\rm s}$. Their fit with a LP gives $\chi^2/{\rm ndf} = 14.1/9 \simeq 1.57$ and $\chi^2/{\rm ndf} = 37.3/10 \simeq 3.73$ for the considered time intervals, where ${\rm ndf}$ is the number of degrees of freedom. Instead, the fitting with a PLEC similarly yields $\chi^2/{\rm ndf} = 10.1/9 \simeq 1.12$ and $\chi^2/{\rm ndf} = 18.3/10 \simeq 1.83$. Because the fit with a PLEC has a  smaller reduced $\chi^2$ the LHAASO collaboration prefers $12.5_{- 2.4}^{+ 3.2} \, {\rm TeV}$.

We adopt $15 \, {\rm TeV}$ as a benchmark energy, representing a conservative intermediate value between the two spectral reconstructions. The adopted $15 \, \rm TeV$ benchmark value should be understood within the assumptions of the intrinsic spectral reconstruction, and alternative spectral models may partially affect the quantitative interpretation. Note that by assuming the two quoted LHAASO energies our results on the required ALP parameter space to justify the LHAASO photons are totally unchanged. We also quote a part of the abstract of the LHAASO paper~\cite{lhaaso2}: ``Observations of gamma-rays up to 13 teraelectronvolts from a source with a measured redshift of $z = 0.151$ hints more transparency in intergalactic space than previously expected. Alternatively, one may invoke new physics such as Lorentz invariance violation or an axion origin of very high-energy signals.''. Still, in this work we show that LIV is unable to justify LHAASO detection within current LIV bounds and the required EBL level proposed by LHAASO is not very physical for two reasons (see Fig. 4 in the LHAASO paper~\cite{lhaaso2}): 1) the EBL density is a discontinuous function of the EBL photon wavelength; 2) it is too close and in some case even weaker than galaxy counts, which represent a lower limit on the EBL.


\

\section{Extragalactic background light (EBL)\label{SecEBL}}

The extragalactic background light (EBL) encompasses the infrared, optical, and ultraviolet emission from the whole stellar population throughout the cosmic history, with additional  reprocession by dust (see~\cite{dwekkrennrich} for a review). The determination of the EBL spectral number density $n_{\rm EBL} \bigl(\epsilon (z), z \bigr)$ at EBL photon energy $\epsilon$ and generic redshift $z$ is subject to significant uncertainties, primarily due to foreground contamination from zodiacal light, which exceeds the EBL by several orders of magnitude~\cite{hauwserdwek2001}. Very-high-energy (VHE) photons emitted by astrophysical sources can interact with EBL photons via the Breit-Wheeler process  
$\gamma_{\rm VHE} + \gamma_{\rm EBL} \to e^+ + e^-$, and may thus be absorbed~\cite{breitwheeler,heitler}. In order to evaluate the EBL optical depth $\tau_{\rm EBL}$ for astrophysical photons produced by a source at redshift $z_s$ with observed energy ${\cal E}_0$, we employ the result first derived 
in~\cite{nishikov1962,gouldschreder1967,faziostecker1970}, which reads
\begin{eqnarray}
&\displaystyle \tau_{\rm EBL} ({\cal E}_0, z_s) = \int_0^{z_s} dz ~ \frac{d l (z)}{d z} \int_{- 1}^1 d ({\rm cos} \, \theta) \, \frac{1 - \, {\rm cos} \, \theta}{2} \int_{\epsilon_{\rm thr} ({\cal E} (z), \theta)}^{\infty} d \epsilon (z) ~  \times   \nonumber \\
&\displaystyle \times \, n_{\rm EBL} \bigl(\epsilon (z), z \bigr) \, \sigma \bigl({\cal E} (z), \epsilon (z), \theta \bigr)    \label{marco10062017a}
\end{eqnarray}
with $\sigma \bigl({\cal E} (z), \epsilon (z), \theta \bigr)$ the Breit-Wheeler cross section~\cite{breitwheeler,heitler} and $\theta$ the scattering angle. Furthermore, in Eq.~(\ref{marco10062017a}) $\epsilon_{\rm thr} ({\cal E} (z), \theta)$ is the threshold 
energy for an emitted photon, given by
\begin{equation}
\epsilon_{\rm thr} ({\cal E} (z), \theta) \equiv \frac{2 m_e^2c^4}{{\cal E}(z)(1 - \cos{\theta})} ~,
\label{ethr}
\end{equation}
where $m_e$ and $c$ are the electron mass and the speed of light, respectively. Finally, the distance covered by the emitted photon per unit redshift at redshift $z$ is 
\begin{equation}
\frac{d l (z)}{d z} = \frac{c}{H_0} \frac{1}{(1 + z) \, \left[\Omega_{\Lambda} + \Omega_M \, (1 + z )^3 \right]^{1/2}}~,
\label{marco10062017b}
\end{equation}
with Hubble constant $H_0 \simeq 70 \, {\rm km} \, {\rm s}^{- 1} \, {\rm Mpc}^{- 1}$ and dimensionless mean cosmic density of matter and dark energy $\Omega_M  \simeq 0.3$ and $\Omega_{\Lambda} \simeq 0.7$, respectively. The most difficult task in evaluating $\tau_{\rm EBL}({\cal E}_0, z_s)$ is the determination of $n_{\rm EBL} \bigl(\epsilon (z), z \bigr)$ since it is affected by the uncertainties stated above. Several different approaches have been developed to find the EBL level: 1) forward evolution~\cite{primack2001,primack2005,gilmore2009,gilmore2012,inoue2013}, 2) backward evolution from the ground~\cite{frv2008,franceschinirodighiero2017}, 3) backward evolution from the sky~\cite{ciber2017,saldanalopez2021}, 4) inferred evolution~\cite{kneiske2002,kneiske2004,finke2010}, 5) minimal EBL~\cite{madaupozzetti2000,kneiskedole2010}, 6) observed evolution~\cite{dominguez2011}, and 7) compared observations~\cite{schroedter2005,aharonian2006,mazinraue2007,mazingoebel2007,
finkerazzaque2009,orrkrennrichdwek2011}. While all previous models yield similar results in the optical and UV ranges, they give rise to different predictions in the IR band, which is particularly important as it is responsible for the strong absorption of TeV photons~\cite{dgr2013}. Since measurements made by satellite borne detectors can effectively remove foreground effects -- especially zodiacal light -- by probing different directions and enabling its accurate subtraction, their results are more reliable. Therefore, we consider the most recent approach of this kind, namely the model of Saldana-Lopez et al. -- referred to as SL EBL model~\cite{saldanalopez2021} -- which has also been employed by the LHAASO collaboration~\cite{lhaaso1}. 

Table I reports the values of the optical depth $\tau_{\rm CP}$ computed according to conventional physics (CP) for several EBL models at the energies ${\cal E}$ relevant to LHAASO ($15 \, \rm TeV$)~\cite{lhaaso1,lhaaso2} and Carpet ($300 \, \rm TeV$)~\cite{carpet3} detections, along with the corresponding photon survival probability 

\begin{equation}
P_{\rm CP} ({\cal E}; \gamma \to \gamma) = e^{- \, \tau_{\rm CP}({\cal E})}~.
\label{19012026a}
\end{equation}

\begin{table}[h]

\label{tabEBL}
\begin{tabular}{l|cc|cc}
\hline
\hline
\multicolumn{1}{c|}{EBL model} &\multicolumn{2}{c|}{$15 \, \rm TeV$} &\multicolumn{2}{c}{$300 \, \rm TeV$} \\

\hline

& $\tau_{\rm CP}$ & $P_{\rm CP}$  & $\tau_{\rm CP}$ & $P_{\rm CP}$ \\
D  &  $12.7$ & $3 \times 10^{-6}$ & $\gg 65$\footnote{Reported $\tau_{\rm CP}$ for model D~\cite{dominguez2011} are up to ${\cal E} =  30 \, \rm TeV$ (see: \url{https://side.iaa.es/EBL/}).} & $\sim 0$ \\
G  & $9.6$ & $6 \times 10^{-5}$ &  $>176$\footnote{Reported $\tau_{\rm CP}$ for model G~\cite{gilmore2012} are up to ${\cal E} = 100 \, \rm TeV$ (see: \url{https://physics-legacy.pbsci.ucsc.edu/~joel/EBLdata-Gilmore2012/}).} & $\sim 0$ \\
FR & $10.1$ & $4 \times 10^{-5}$ &  $24881$ & $\sim 0$ \\
SL &  $13.4$ & $2 \times 10^{-6}$ & $>222$\footnote{Reported $\tau_{\rm CP}$ for model SL~\cite{saldanalopez2021} are up to ${\cal E} = 100 \, \rm TeV$ (see: \url{https://www.ucm.es/blazars/ebl/}).} & $\sim 0$ \\

\hline
\hline
\end{tabular}
\caption{Values of the optical depth $\tau_{\rm CP}$ and corresponding photon survival probability $P_{\rm CP}$ at the energies of LHAASO ($15 \, \rm TeV$)~\cite{lhaaso1,lhaaso2} and Carpet ($300 \, \rm TeV$)~\cite{carpet3} detections for the EBL models of Dom\'inguez et al. (D)~\cite{dominguez2011}, Gilmore et al. (G)~\cite{gilmore2012}, Franceschini and Rodighiero (FR),~\cite{franceschinirodighiero2017} and Saldana-Lopez et al. (SL)~\cite{saldanalopez2021}.}
\end{table}

Using the last line of TABLE I for ${\cal E} = 300 \, {\rm TeV}$ and Eq. (1) of the main text we find that the number of expected events observed by Carpet according to conventional physics is indeed $N^{\rm CP}_{\gamma} \sim 10^{- 96}$.

\

\section{Axion-like particles (ALPs): Formalism}

This section provides a self-contained overview of the aspects of ALPs that are relevant to the main text. Henceforth, dimensionless units are employed. ALPs -- denoted by $a$ -- are spin-0, neutral, pseudoscalar bosons of mass $m_a$. Their interaction with photons is described by the Lagrangian
\begin{equation}
\label{lagr}
{\cal L}_{\rm ALP} =  \frac{1}{2} \, \partial^{\mu} a \, \partial_{\mu} a - \frac{1}{2} \, m_a^2 \, a^2 - \, \frac{1}{4 } g_{a\gamma\gamma} \, F_{\mu\nu} \tilde{F}^{\mu\nu} a = \frac{1}{2} \, \partial^{\mu} a \, \partial_{\mu} a - \frac{1}{2} \, m_a^2 \, a^2 + g_{a \gamma \gamma} \, 
{\bf E} \cdot {\bf B}~a~,
\end{equation}
where $F_{\mu\nu}$ is the electromagnetic tensor with dual $\tilde{F}^{\mu\nu}$ and with electric and magnetic components ${\bf E}$ and ${\bf B}$, respectively. Throughout our whole discussion ${\bf B}$ is an {\it external} magnetic field whereas ${\bf E}$ is the electric field of a propagating photon. As it is evident from the ${\bf E} \cdot {\bf B} \, a$ term in Eq. (\ref{lagr}), the ALP field $a$ couples only to the transverse component ${\bf B}_T$ of ${\bf B}$, which is perpendicular to ${\bf E}$ (see also~\cite{dgr2011}). We stress that the ALP mass $m_a$ and $g_{a \gamma \gamma}$ are unrelated parameters. 

Whenever the external magnetic field is rather strong, photon one-loop vacuum polarization effects become relevant and are described by the Heisenberg-Euler-Weisskopf (HEW) effective Lagrangian
\begin{equation}
\label{HEW}
{\cal L}_{\rm HEW} = \frac{2 \alpha^2}{45 m_e^4} \, \left[ \left({\bf E}^2 - {\bf B}^2 \right)^2 + 7 \left({\bf E} \cdot {\bf B} \right)^2 \right] 
\end{equation}
with $\alpha$ the fine-structure constant and $m_e$ the electron mass~\cite{hew1, hew2, hew3}. 

Starting form Eq.~(\ref{lagr}), a photon-ALP beam of energy $\cal E$ propagating along $y$-direction in an external magnetic field $\bf B$ and in the regime ${\cal E} \gg m_a$ -- a condition always met in our discussion -- is described by the Schr\"odinger-like equation~\cite{rs}
\begin{equation}
\label{propeq} 
\left(i \, \frac{d}{d y} + {\cal E} +  {\cal M} ({\cal E},y) \right)  \psi(y)= 0~,
\end{equation}
where ${\cal M} ({\cal E},y)$ is the mixing matrix, while $\psi(y)$ is the wave function reading
\begin{equation}
\label{psi} 
\psi(y)=\left(\begin{array}{c}A_x (y) \\ A_z (y) \\ a (y) \end{array}\right)~,
\end{equation}
where $A_x (y)$ and $A_z (y)$ stand for the photon linear polarization amplitudes along the $x$- and $z$-axis, respectively, while $a (y)$ is the ALP amplitude. Moreover, denoting as 
$\phi$ the angle between ${\bf B}_T$ and the $z$-axis, ${\cal M} ( {\cal E}, y)$ reads 
\begin{equation}
\label{mixmat}
{\cal M} ({\cal E},y) \equiv \displaystyle \left(
\begin{array}{ccc}
\Delta_{xx} ({\cal E},y) & \Delta_{xz} ({\cal E},y) & \Delta_{a \gamma}(y) \, {\rm sin} \, \phi \\
\Delta_{zx} ({\cal E},y) & \Delta_{zz} ({\cal E},y) & \Delta_{a \gamma}(y) \, {\rm cos} \, \phi \\
\Delta_{a \gamma}(y) \, {\rm sin}  \, \phi & \Delta_{ a \gamma}(y) \, {\rm cos} \, \phi & \Delta_{a a} ({\cal E}) \\
\end{array}
\right)~,
\end{equation}
where
\begin{equation}
\label{deltaxx}
\Delta_{xx} ({\cal E},y) \equiv \Delta_{\bot} ({\cal E},y) \, {\rm cos}^2 \, \phi + \Delta_{\parallel} ({\cal E},y) \, {\rm sin}^2 \, \phi~,
\end{equation}
\begin{equation}
\label{deltaxz}
\Delta_{xz} ({\cal E},y) = \Delta_{zx} ({\cal E},y) \equiv \left(\Delta_{\parallel} ({\cal E},y) - \Delta_{\bot} ({\cal E},y) \right) {\rm sin} \, \phi \, {\rm cos} \, \phi~,
\end{equation}
\begin{equation}
\label{deltazz}
\Delta_{zz} ({\cal E},y) \equiv \Delta_{\bot} ({\cal E},y) \, {\rm sin}^2 \, \phi + \Delta_{\parallel} ({\cal E},y) \, {\rm cos}^2 \, \phi~,
\end{equation}
\begin{equation}
\label{deltamix} 
\Delta_{a \gamma}(y) = \frac{1}{2}g_{a\gamma\gamma}B_T(y)~,
\end{equation}
\begin{equation}
\label{deltaM} 
\Delta_{aa} ({\cal E}) = - \frac{m_a^2}{2 {\cal E}}~,
\end{equation}
and
\begin{equation}
\Delta_{\bot} ({\cal E},y) = \frac{i}{2 \, \lambda_{\gamma} ({\cal E},y)} - \frac{\omega^2_{\rm pl}(y)}{2 {\cal E}} + \frac{2 \alpha}{45 \pi} \left(\frac{B_T(y)}{B_{{\rm cr}}} \right)^2 {\cal E} + \rho_{\rm CMB} \, {\cal E}~, 
\label{deltaort} 
\end{equation}
\begin{equation}
\Delta_{\parallel} ({\cal E},y) = \frac{i}{2 \, \lambda_{\gamma} ({\cal E},y)} - \frac{\omega^2_{\rm pl}(y)}{2 {\cal E}} + \frac{7 \alpha}{90 \pi} \left(\frac{B_T(y)}{B_{{\rm cr}}} \right)^2 {\cal E} + \rho_{\rm CMB} \, {\cal E} ~,  
\label{deltapar} 
\end{equation}
with $B_{{\rm cr}} \simeq 4.41 \times 10^{13} \, {\rm G}$ the critical magnetic field and 
$\rho_{\rm CMB}=0.522 \times 10^{-42}$. Eq.~(\ref{deltamix}) encodes the interaction between photons and ALPs, while Eq.~(\ref{deltaM}) accounts for the contribution of the ALP mass. The first terms in Eqs.~(\ref{deltaort}) and~(\ref{deltapar}) represent the photon absorption by the EBL and $\lambda_{\gamma}$ denotes the $\gamma + \gamma \to e^+ + e^-$ mean free path. The second term in the same equations involves the plasma frequency $\omega_{\rm pl}$ which is determined by the electron number density $n_e$ through the relation $\omega_{\rm pl}=(4 \pi \alpha n_e / m_e)^{1/2}$. The third terms express the QED vacuum polarization arising from ${\cal L}_{\rm HEW}$ in Eq.~(\ref{HEW}), which induces polarization-dependent effects and birefringence. Finally, the last terms describe the photon dispersion on the cosmic microwave background (CMB)~\cite{raffelteffect}, an effect which becomes important over extragalactic distances~\cite{grjhea}.

A remark is important in view of our subsequent developments. If the photon had a nonvanishing mass $m_{\gamma}$ then the following replacements should be made in 
Eqs. (\ref{deltaort}) and (\ref{deltapar})
\begin{equation} 
\Delta_{\parallel} ({\cal E},y) \to \Delta_{\parallel} ({\cal E},y) - \, \frac{m^2_{\gamma}}{2 {\cal E}}~, \ \ \ \ \ \ \Delta_{\bot} ({\cal E},y) \to \Delta_{\bot} ({\cal E},y) - \, \frac{m^2_{\gamma}}{2 {\cal E}}~.
\label{29012026q}
\end{equation}

It is worth noting that in the regime ${\cal E} \gg m_a$ the relativistic photon-ALP system can be formally treated as a three-level, non-relativistic, unstable quantum system. The off-diagonal structure of the mixing matrix (\ref{mixmat}) implies that the propagation eigenstates differ from the interaction eigenstates, leading to $\gamma \leftrightarrow a$ {\it oscillations} analogous to flavor oscillations of massive neutrinos~\cite{mpz}. The key-difference is that here an external magnetic field $\bf B$ is required in order to compensate for the spin mismatch between photons and ALPs.

We denote by ${\cal U}({\cal E};y,y_0)$ the {\it transfer matrix} associated with Eq. (\ref{propeq}) and defined as the particular solution satisfying the initial condition ${\cal U}({\cal E};y_0,y_0)=1$. The general solution of Eq. (\ref{propeq}) can then be written as
\begin{equation}
\label{psi2} 
\psi(y)={\cal U}({\cal E};y,y_0)\psi(y_0)~.
\end{equation}

Since the polarization of photons at the energies considered here cannot be measured, the best we can do is to assume that they are unpolarized. As a consequence, they must be described by means of the generalized polarization density matrix $\rho(y)$. Accordingly, 
$\rho(y)$ satisfies the Liouville-von Neumann equation associated with Eq.~(\ref{propeq})
\begin{equation}
\label{vneum}
i \frac{d \rho (y)}{d y} = \rho (y) \, {\cal M}^{\dag} ( {\cal E}, y) - {\cal M} ( {\cal E}, y) \, \rho (y)~,
\end{equation}
whose solutions can be represented by means of ${\cal U} \bigl( {\cal E}; y, y_0 \bigr)$ as
\begin{equation}
\label{unptrmatr}
\rho ( y ) = {\cal U} \bigl({\cal E}; y, y_0 \bigr) \, \rho_0 \, {\cal U}^{\dag} \bigl({\cal E}; y, y_0 \bigr)~.
\end{equation}
Just like in quantum mechanics, the probability that the beam evolves from the initial state 
$\rho_0$ at position $y_0$ to the final state $\rho$ at position $y$ is 
\begin{equation}
\label{unpprob}
P_{\rho_0 \to \rho} ({\cal E},y) = {\rm Tr} \Bigl[\rho \, {\cal U} ({\cal E}; y, y_0) \, \rho_0 \, {\cal U}^{\dag} ({\cal E}; y, y_0) \Bigr]  
\end{equation}
with ${\rm Tr} \, \rho_0 = {\rm Tr} \, \rho =1$~\cite{dgr2011}.

To clarify the properties of the different regimes defined by the relative importance of the various $\Delta$ terms in Eq.~(\ref{mixmat}), we consider a simplified scenario for illustrative purposes. Specifically, we assume that: 1) photons are fully polarized, 2) absorption is negligible (i.e., $\lambda_{\gamma} \to \infty$), 3) the medium is homogeneous, and 4) the external magnetic field $\bf B$ is uniform, i.e. ${\bf B}(y) \equiv {\bf B}$. The last condition allows us to choose the $z$-axis along the direction of 
${\bf B}_T$, which corresponds to set $\phi = 0$ in Eq.~(\ref{mixmat}). Under these assumptions the photon-to-ALP conversion probability takes the form 
\begin{equation}
\label{convprob}
P_{\gamma \to a} ({\cal E}, y) = \left(\frac{g_{a\gamma\gamma}B_T \, l_{\rm osc} ({\cal E})}{2\pi} \right)^2 {\rm sin}^2 \left(\frac{\pi (y-y_0)}{l_{\rm osc} ({\cal E})} \right)~,
\end{equation}
where
\begin{equation}
\label{losc}     
l_{\rm osc} ({\cal E}) \equiv \frac{2 \pi}{\left[\bigl(\Delta_{zz} ({\cal E}) - \Delta_{aa} ({\cal E}) \bigr)^2 + 4 \, \Delta_{a\gamma}^2 \right]^{1/2}}~
\end{equation}
is the photon-ALP {\it oscillation length}. It proves useful to define the {\it low-energy threshold} ${\cal E}_L$ and the {\it high-energy threshold} ${\cal E}_H$ as 
\begin{equation}
\label{EL}
{\cal E}_L \equiv \frac{|m_a^2 - \omega^2_{\rm pl}|}{2 g_{a \gamma \gamma} \, B_T}   
\end{equation}
and 
\begin{equation}
\label{EH}
{\cal E}_H \equiv g_{a \gamma \gamma} \, B_T \left[\frac{7 \alpha}{90 \pi} \left(\frac{B_T}{B_{\rm cr}} \right)^2 + \rho_{\rm CMB} \right]^{- 1}~,
\end{equation} 
respectively. For energies in the range ${\cal E}_L \lesssim {\cal E} \lesssim {\cal E}_H$ the system is in the {\it strong-mixing} regime, where the effects of plasma, ALP mass, QED one-loop corrections, and photon dispersion on the CMB become negligible with respect to the photon-ALP mixing term in Eq.~(\ref{deltamix}). Therefore $P_{\gamma \to a} (y)$ becomes maximal and energy independent. It reads
\begin{equation}
\label{convprobSM}
P_{\gamma \to a} (y) = {\rm sin}^2 \left( \frac{g_{a\gamma\gamma} B_T}{2}  (y-y_0) \right)~.
\end{equation}
For energies ${\cal E} \lesssim {\cal E}_L$ the plasma and/or the ALP mass contributions dominate over the other terms in Eq.~(\ref{mixmat}). Similarly, in the range ${\cal E} \gtrsim {\cal E}_H$ the QED one-loop vacuum polarization effects and/or the photon dispersion on the CMB become dominant. In either case, the system is in the {\it weak-mixing} regime, where $P_{\gamma \to a} ({\cal E}, y)$ becomes energy-dependent and progressively decreasing.

The above simplified scenario can be extended to the general case, albeit with considerably larger analytical complexity. At any rate, in this Letter we compute the exact expression for photon-ALP propagation including an accurate description of the magnetization, dispersion, and absorption properties of the media crossed by the photon-ALP beam.
 
\

\section{Bounds on the ALP parameter space}

The only laboratory upper bound on $g_{a\gamma\gamma}$ is provided by the CAST experiment and reads $g_{a\gamma\gamma} < 0.66 \times 10^{-10} \, {\rm GeV}^{-1}$ for $m_a < 0.02 \, {\rm eV}$ at $95 \%$ CL~\cite{cast}.

Additional constraints on the ALP parameter space have been derived from the absence of ALP-induced signatures in various astrophysical observations~\cite{straniero,payez2015,fermi2016,berg,conlonLim,limFabian,limJulia,limKripp,limRey2,mwd}. Notably, a bound identical to that of CAST has independently emerged from studies of stellar evolution in globular clusters~\cite{straniero}. The bound  $g_{a\gamma\gamma} < 5.3 \times 10^{-12} \, {\rm GeV}^{-1}$ for $m_a < 4.4 \times 10^{-10} \, {\rm eV}$ has been inferred from the non-detection of photon reconversion from ALPs produced in the supernova SN1987A but no confidence level has been specified~\cite{payez2015}. This result has subsequently been challenged~\cite{barblumdamico2020}.

The tightest constraints currently available are the following. One originates from polarization measurements of radiation emitted by magnetic white dwarfs (MWDs), yielding $g_{a\gamma\gamma} < 5.4 \times 10^{-12} \, {\rm GeV}^{-1}$ for $m_a < 3 \times 10^{-7} \, {\rm eV}$ at $95 \%$ CL, even though it may turn out to be too strong owing to foreground effects~\cite{mwd}. Further tight constraints apply to very light ALPs with mass $m_a < \mathcal{O}(10^{-12}) \, {\rm eV}$ based on X-ray observations of blazars and/or galaxy clusters, typically excluding values of $g_{a\gamma\gamma} > \mathcal{O}(10^{-12}) \, {\rm GeV}^{-1}$ at $95 \%$ CL~\cite{berg,conlonLim,limFabian,limJulia,limKripp,limRey2}.

\

\section{Photon-ALP beam propagation in different regions}

In this Section our goal is to investigate how the photon-ALP beam propagates from its source inside GRB 221009A to us, crossing the GRB itself, the host galaxy, the extragalactic space, and the Milky Way. Below, we outline the key-properties of each of these environments which affect the beam evolution. 

\subsection{Gamma-ray burst (GRB)}

For GRB 221009A we adopt the standard afterglow scenario in which multi-TeV photons are generated in the external forward shock driven by the GRB jet into the surrounding medium, as observed in other GRBs detected at TeV energies. Before escaping from the source, these photons propagate through the downstream region, whose comoving length is $\Delta R^{\prime} \sim R^{\prime}/\Gamma$ with $R^{\prime}$ and $\Gamma$ denoting the distance from the central engine and the bulk Lorentz factor, respectively~\cite{bm76}.

We consider both a constant-density interstellar medium and a wind-like environment shaped by the progenitor star. The comoving magnetic field $B^{\prime}$ is derived from the shock conditions, assuming that a fraction $\epsilon_B$ of the shock-dissipated energy is transferred to the magnetic field. The key-parameter governing the photon-ALP conversion in the source is $B^{\prime} R^{\prime}/\Gamma$, which reaches its maximum during the deceleration phase of the blast wave. For physically realistic values of the GRB parameters -- e.g. jet kinetic energy ${\cal E}_k = 10^{54} \, \rm erg$, constant medium density $n_0 = 10 \, \rm cm^{-3}$, $\epsilon_B = 10^{-3}$ -- the quantities at maximal luminosity are $B^{\prime} \simeq 2 \rm G$, $R^{\prime} \simeq 2 \times 10^{17} \, \rm cm$ and $\Gamma \simeq 45$, yielding a comoving propagation length $\Delta R^{\prime} \simeq 5 \times 10^{15} \, \rm cm$~\cite{navasironi,derishevpiran,derishev}. 

We compute the transfer matrix ${\cal U}_1 ({\cal E}; y_2, y_1)$ describing the photon-ALP beam propagation inside the GRB, where $y_1$ and $y_2$ denote the positions of the photon production site and the outer edge of the GRB, respectively. Following the method considered in~\cite{gtre2019}, we find that the propagation length is too short for significant photon-ALP conversion under the above conditions, and thus ${\cal U}_1 ({\cal E}; y_2, y_1) \simeq 1$ (for more details see~\cite{noi}).

\subsection{Host galaxy}

As we have seen in Section I, GRB 221009A is close to the center of a spiral galaxy 
viewed edge-on in first approximation. Therefore, the VHE photon beam lies inside the galactic disk of the host. This fact drastically simplifies our job since we need to know only the relevant quantities in the disk. They are the electron density -- which fixes the local plasma frequency -- and the magnetic field. According to the standard lore a realistic value for the electron number density is $n_{{\rm host}, \, e} = 1 \, \rm cm^{-3}$~\cite{SpiralBrev}. 
In general, the magnetic field of spirals consists of three components: 1) a regular field 
${\bf B}_{\rm reg}$, 2) an anisotropic turbulent component at large scale ${\bf B}_{\rm aniso}$, and 3) an isotropic turbulent component at small scale ${\bf B}_{\rm iso}$. Each of them  is treated independently (for a review see~\cite{SpiralBrev}). Their radial dependence along the $y$-direction from the host center to us is modeled using an exponential profile~\cite{Heesen2023}, and its coherence properties are described by a Kolmogorov-type turbulence power spectrum $M(k)\propto k^q$ with index $q = -11/3$ and wave number $k$ ranging between $k_L = 2\pi/\Lambda_{\rm max}$ and $k_H = 2\pi/\Lambda_{\rm min}$. The general form used for each component ${\bf B}$ of ${\bf B}_{\rm gal}$ is 
\begin{equation}     
\label{Bprofile}
B(y) = {\cal B} \left(B_0,k,q,y \right) \exp{(-y/r_0)}~,
\end{equation}
where ${\cal B}$ encodes the turbulence spectrum, while $B_0$ and $r_0$ denote the central field strength and radial scale length, respectively~\cite{meyerKolm}. In the disk the situation simplifies considerably since the regular component dominates, and so only ${\bf B}_{\rm reg}$ has to be considered. Still, it is often assumed that ${\bf B}_{\rm reg}$ is smooth~\cite{Fletcher2010} but since the leading contribution to the magnetic field of spirals arises from the existence of spiral arms and from star formation we adopt a more physically motivated modeling for ${\bf B}_{\rm reg}$ using a Kolmogorov spectrum, with coherence scales between $\Lambda_{\rm min}$ and $\Lambda_{\rm max}$. Moreover, in Eq.~(\ref{Bprofile}) a realistic value is $r_0 = 15 \, \rm kpc$ (radial scale length). Finally, in order to figure out the effect of $B_{\rm reg}$ on the photon-ALP conversion probability we shall consider two cases for the central field strength. 
\begin{enumerate} 
\item Standard host magnetic field: $B_0 = 20 \, \mu{\rm G}$ with coherence scales between $\Lambda_{\rm min}= 2 \, \rm kpc$ and $\Lambda_{\rm max}= 4 \, \rm kpc$. 
\item High host magnetic field: $B_0 = 50 \, \mu{\rm G}$ with coherence scales between $\Lambda_{\rm min}= 1.5 \, \rm kpc$ and $\Lambda_{\rm max}= 3 \, \rm kpc$. 
\end{enumerate}
The first choice is typical of a generic spiral -- note that the Milky Way has an anomalously low value of $B_0$ -- while the second one is motivated by the fact that the host shows a more-than-average star formation rate. While both values are consistent with observations of edge-on spiral galaxies~\cite{SpiralBrev,Heesen2023}, our choice has the advantage to show how the photon-ALP conversion efficiency depends on the host magnetic field. It is then straightforward to evaluate the transfer matrix describing photon-ALP conversion within the host galaxy, denoted by ${\cal U}_2 ({\cal E}; y_3, y_2)$ where $y_3$ is the position of the outer luminous galaxy radius which is assumed to be $y_3 = 20 \, \rm kpc$.

\subsection{Extragalactic space}

Our knowledge of the extragalactic magnetic field ${\bf B}_{\rm ext}$ remains nowadays very poor. A rather clear-cut upper bound is $B_{\rm ext} < 1.7 \times 10^{- 9} \, {\rm G}$ with a coherence length ${\cal O} (1) \, {\rm Mpc}$~\cite{upbbext}. Unfortunately, the derivation of a lower bound depends on several unknowns, such as the coherence length, the variability of the measured sources of VHE gamma rays, the time span over which they are monitored, their duty cycle and the details of the induced cascades (a clear review of this topic is~\cite{alvesbatista2021}). Hence a large spread in the results exists. As an example of the resulting uncertainty, we quote the most recent lower bound: $B_{\rm ext} > 7.1 \times 10^{- 16} \, {\rm G}$ for a coherence length of $1 \, {\rm Mpc}$ and a source duty cycle of $10 \, {\rm yr}$, which becomes $B_{\rm ext} > 1.8 \times 10^{- 14} \, {\rm G}$ and $B_{\rm ext} > 3.9 \times 10^{- 14} \, {\rm G}$ for a duty cycle of $10^4 \, {\rm yr}$ and $10^7 \, {\rm yr}$, respectively~\cite{aharonian2023}. 

While the possibility of a very small extragalactic magnetic field -- say $B_{\rm ext} < 
10^{- 15} \,  {\rm G}$ -- cannot be excluded, since about twenty years ${\bf B}_{\rm ext}$ has been described by a very specific model. It consists of a domain-like network, where ${\bf B}_{\rm ext}$ is assumed homogeneous over a whole domain of size $L_{\rm dom}$ equal to its coherence length, with ${\bf B}_{\rm ext}$ changing randomly its direction from one domain to the next keeping approximately the same strength. Therefore, the photon-ALP beam propagation becomes a {\it random process} and only a single realization at once can be observed. Additionally, it has been assumed that such a change of direction is abrupt, since correspondingly the beam propagation equation (\ref{propeq}) becomes easy to solve~\cite{kronberg1994,grassorubinstein2001}. This scenario -- named {\it domain-like sharp-edges} (DLSHE) -- is physically motivated by the outflows from primeval galaxies further amplified by turbulence~\cite{reessetti1968,hoyle1969,kronbergleschhopp1999,furlanettoloeb2001}. Benchmark values are $B_{\rm ext} = {\cal O} (10^{- 9}) \, {\rm G}$ on a coherence length ${\cal O} (1) \, {\rm Mpc}$ which sets the value of $L_{\rm dom}$ (for more details  see~\cite{galantironcadelli20118prd}). Still, the abrupt change in direction at the interface between two adjacent domains leads to a failure of the DLSHE model at the energies considered here. A way out of this difficulty is to smooth out the sharp edges of the 
domains so that the components of ${\bf B}_{\rm ext}$ change continuously across the interface, thereby leading to the {\it domain-like smooth-edges} (DLSME) model built up in~\cite{galantironcadelli20118prd,kartavtsev}. Only the ALP scenario described in~\cite{galantironcadelli20118prd,kartavtsev} contemplates photon-ALP oscillations in extragalactic space within the DLSME model. Here we work within such a model and we take $B_{\rm ext} \simeq 10^{- 9} \, {\rm G}$ and $L_{\rm dom}$ in the range $(0.2 - 10) \, {\rm Mpc}$ with $\langle L_{\rm dom} \rangle = 2 \, {\rm Mpc}$~\cite{grjhea}. It turns out that above energies of some TeV photon dispersion on the CMB~\cite{raffelteffect} -- see 
Eqs. (\ref{deltaort}) and (\ref{deltapar}) -- makes the probability for photon-ALP oscillations vanishingly small~\cite{grjhea}. 

Given the intrinsic uncertainty of the strength of ${\bf B}_{\rm ext}$, we consider two cases.  
\begin{enumerate} 
\item Effectively vanishing magnetic field: $B_{\rm ext} < 10^{- 15} \,  {\rm G}$.
\item Sizable magnetic field as described above: $B_{\rm ext} = 10^{- 9} \,  {\rm G}$.
\end{enumerate}
Accordingly ${\cal U}_3 ({\cal E}; y_4, y_3)$ is computed, where $y_4$ is the position of the outer luminous edge of the Milky Way.

\subsection{Milky Way}

The propagation of the photon-ALP beam through the Milky Way depends on the structure of the Galactic magnetic field ${\bf B}_{\rm MW}$ as well as on the electron number density $n_{{\rm MW}, \, e}$ which fixes the local plasma frequency $\omega_{{\rm MW, \, pl}}$.

In order to model ${\bf B}_{\rm MW}$ we adopt the framework proposed by Jansson and Farrar~\cite{jansonfarrar1,jansonfarrar2,BMWturb} which includes the disc and halo components, as well as an X-shaped poloidal field at the Galactic center. While this model incorporates both regular and turbulent magnetic components, only the regular field contributes significantly to the photon-ALP conversion in this context since it has been shown that the inclusion of the turbulent component can change the photon survival probability in the Milky Way by at most a factor of 2~\cite{carenza}. We have also cheked the stability of our results against the alternative model by Pshirkov et al.~\cite{pshirkov2011} finding negligible differences.

As far as the electron density is concerned, the most reliable estimate is $n_{{\rm MW}, \, e} \simeq 1.1 \times 10^{-2} \, \rm cm^{-3}$~\cite{yaomanchesterwang 2017} leading to the  plasma frequency $\omega_{{\rm MW, \, pl}} \simeq 3.9 \times 10^{-12} \, \rm eV$. Based on these parameters, the transfer matrix ${\cal U}_4({\cal E}; y_5, y_4)$ -- with $y_5$ denoting the Earth position -- can be computed following the procedure outlined in~\cite{gtre2019}.

\subsection{Total photon-ALP beam propagation}

According to quantum mechanics, the total transfer matrix of the photon-ALP system from the source to the Earth reads
\begin{equation}
{\cal U} ({\cal E}; y_5, y_1) = \prod_{i = 1}^4 {\cal U}_i ({\cal E}; y_{i + 1}, y_i)~,
\label{14022023a}
\end{equation}
while the total photon survival probability in the presence of photon-ALP oscillations  becomes 
\begin{equation}
P_{\rm ALP} ({\cal E}; \gamma \to \gamma) = \sum_{i = x, z} {\rm Tr} \bigl[\rho_i \, {\cal U} ({\cal E}; y_5, y_1) \rho_{\rm unp} \, {\cal U}^{\dagger} ({\cal E}; y_5, y_1) \bigr]~,
\label{14022023b}
\end{equation}
where
\begin{equation}
\label{densphot}
{\rho}_x = \left(
\begin{array}{ccc}
1 & 0 & 0 \\
0 & 0 & 0 \\
0 & 0 & 0 \\
\end{array}
\right)~, \,\,\,\,\,\,\,\,
{\rho}_z = \left(
\begin{array}{ccc}
0 & 0 & 0 \\
0 & 1 & 0 \\
0 & 0 & 0 \\
\end{array}
\right)~,
\end{equation}
represent the pure photon states with polarization along the $x$ and $z$ directions, respectively, and
\begin{equation}
\label{densunpol}
{\rho}_{\rm unp} = \frac{1}{2} \left(
\begin{array}{ccc}
1 & 0 & 0 \\
0 & 1 & 0 \\
0 & 0 & 0 \\
\end{array}
\right)~,
\end{equation}
describes initially unpolarized photons.

\

\section{Results for the photon-ALP system}

We report our results for the photon survival probability in the various cases considered above.

\subsection{Standard host magnetic field}

\begin{figure*}[h]
\begin{center}
\includegraphics[width=.48\textwidth]{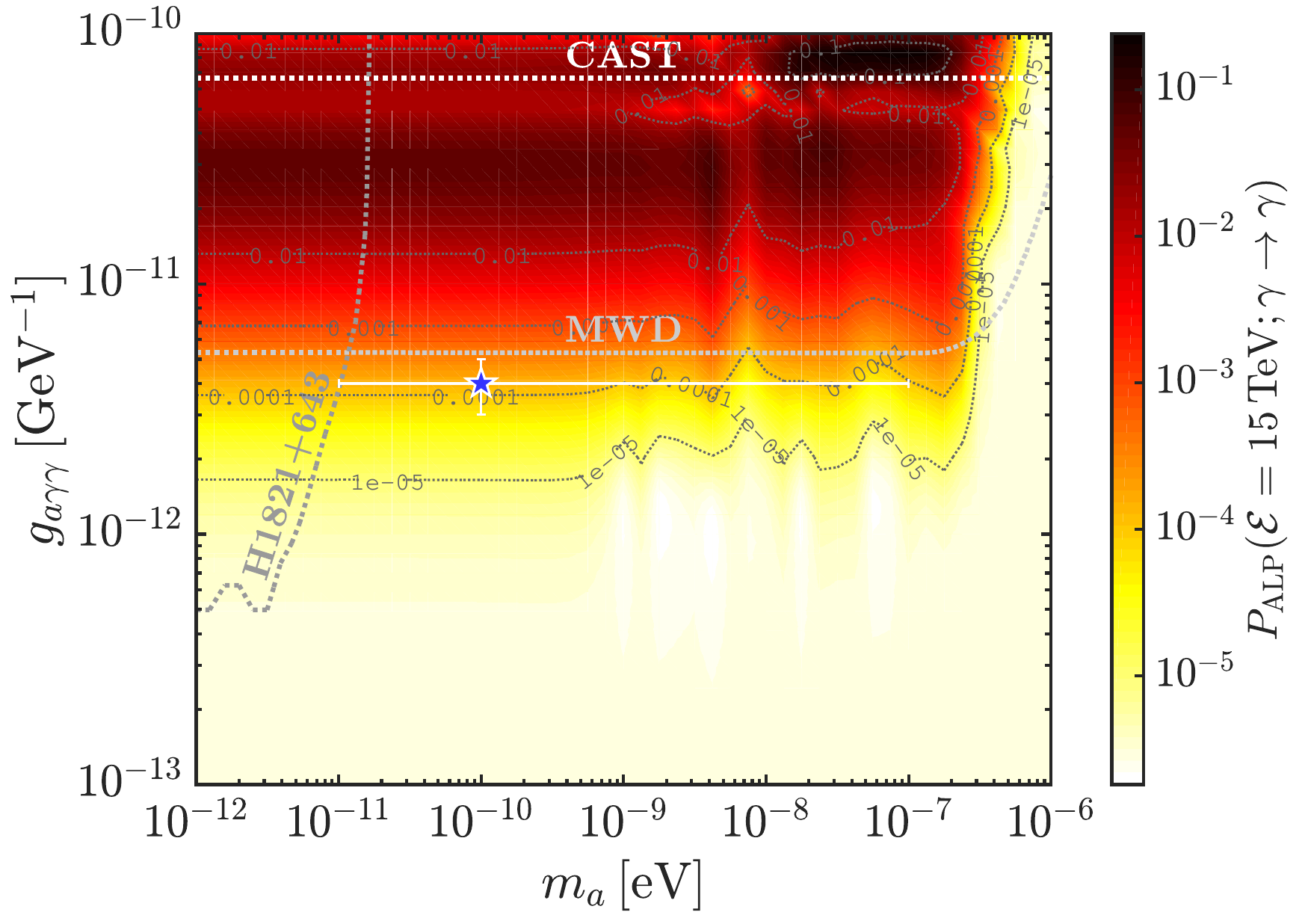} \hspace{3mm} \includegraphics[width=.48\textwidth]{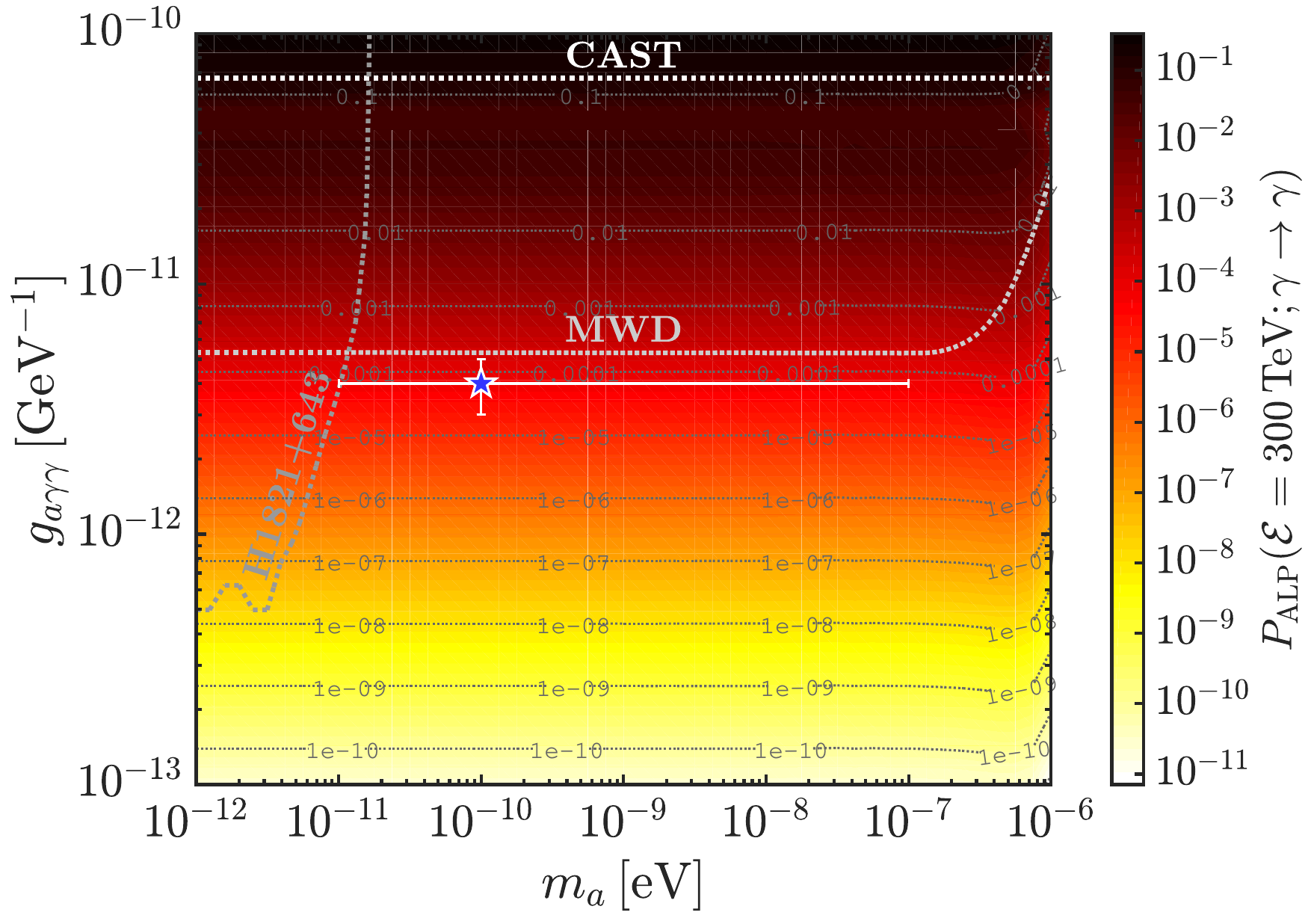}
\end{center}
\caption{\label{parSpaceSpiralBig} Photon survival probability $P_{\rm ALP}({\cal E}; \gamma \to \gamma)$ for energies ${\cal E} = 15 \, \rm TeV$ (left panel) and ${\cal E} = 300 \, \rm TeV$ (right panel), as a function of the ALP mass $m_a$ and the photon-ALP coupling $g_{a\gamma\gamma}$. We assume the SL EBL model, $B_{\rm ext}=10^{- 9} \, \rm G$ and a standard spiral host galaxy. The blue star with error bar marks the choice for the ALP parameters: $m_a=10^{-10} \, \rm eV$ and $g_{a\gamma\gamma}=4 \times 10^{-12} \, \rm GeV^{-1}$. We also show the CAST bound~\cite{cast}, that arising from magnetic white dwarf (MWD) polarization~\cite{mwd} and the one obtained from H1821+643~\cite{limJulia}.}
\end{figure*}      

\begin{figure}[H]
\begin{center}
\includegraphics[width=.45\textwidth]{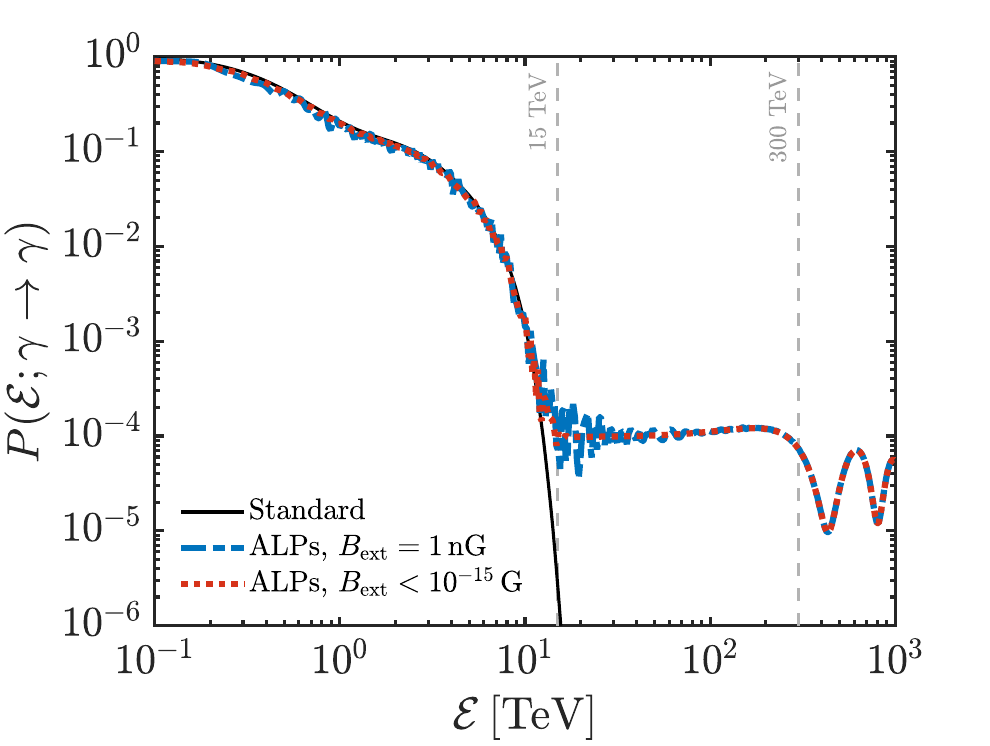}
\end{center}
\caption{\label{survProbFigSpiral} Photon survival probability $P ({\cal E}; \gamma \to \gamma)$ as a function of energy ${\cal E}$ within conventional physics and including the photon-ALP oscillations (for $B_{\rm ext} = 10^{- 9} \, \rm G$ and $B_{\rm ext} < 10^{-15} \, \rm G$) in the case of a standard spiral host galaxy and adopting the SL EBL model. We set the ALP parameters to $m_a=10^{-10} \, \rm eV$ and $g_{a\gamma\gamma}=4 \times 10^{-12} \, \rm GeV^{-1}$.}
\end{figure}

\begin{figure}[H]
\begin{center}
\includegraphics[width=.75\textwidth]{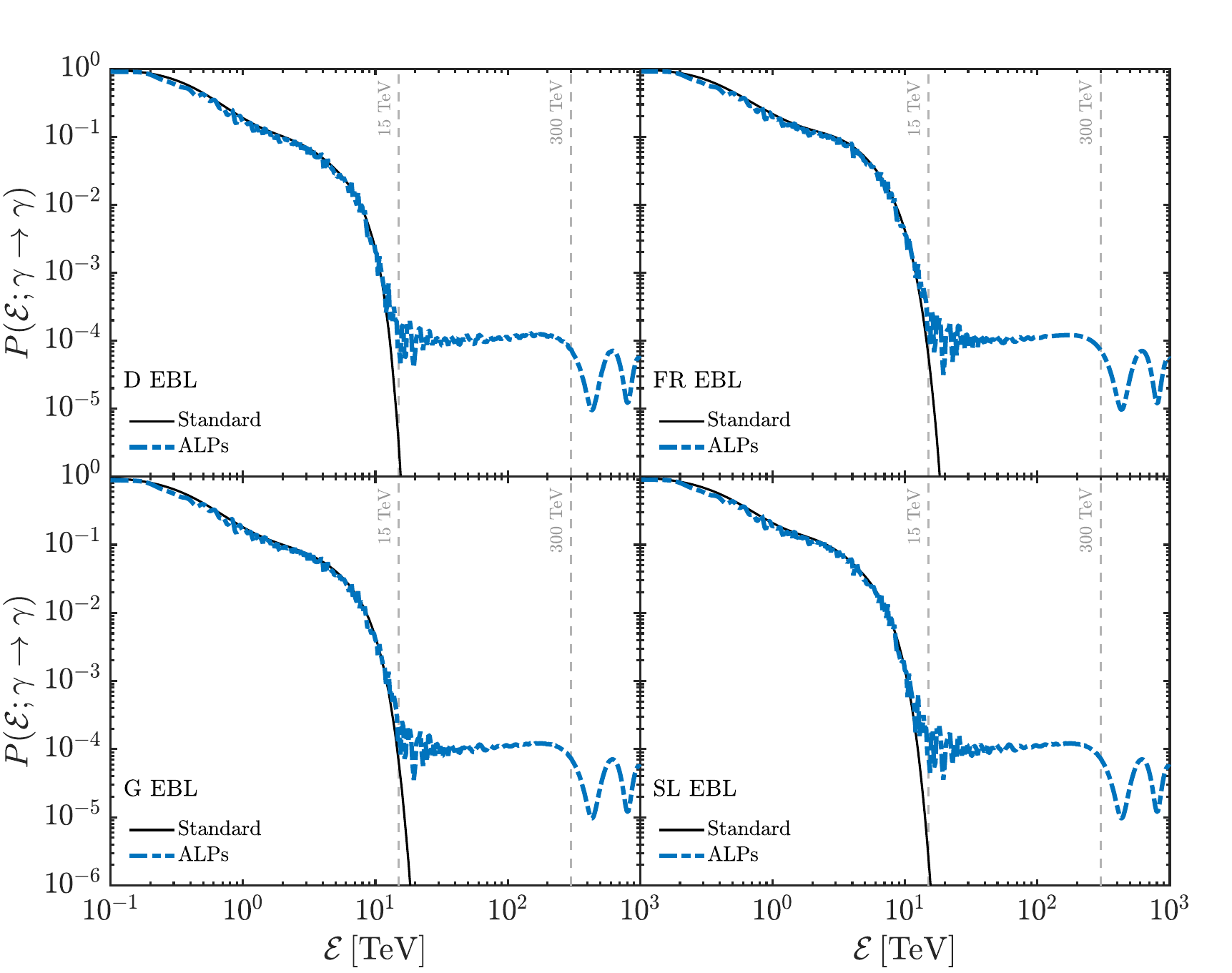}
\end{center}
\caption{\label{survProbSpiralALPsEBLpanelFIG} Photon survival probability $P ({\cal E}; \gamma \to \gamma)$ as a function of energy ${\cal E}$ within conventional physics and including photon-ALP oscillations (for $B_{\rm ext} = 10^{- 9} \, \rm G$) in the case of a standard spiral host galaxy and adopting four different EBL models: 1) D EBL (upper left panel), 2) G EBL (lower left panel), 3) FR EBL (upper right panel), 4) SL EBL (lower right panel). We set the ALP parameters to $m_a=10^{-10} \, \rm eV$ and $g_{a\gamma\gamma}=4 \times 10^{-12} \, \rm GeV^{-1}$.}
\end{figure} 

\begin{figure}[H]
\begin{center}
\includegraphics[width=.45\textwidth]{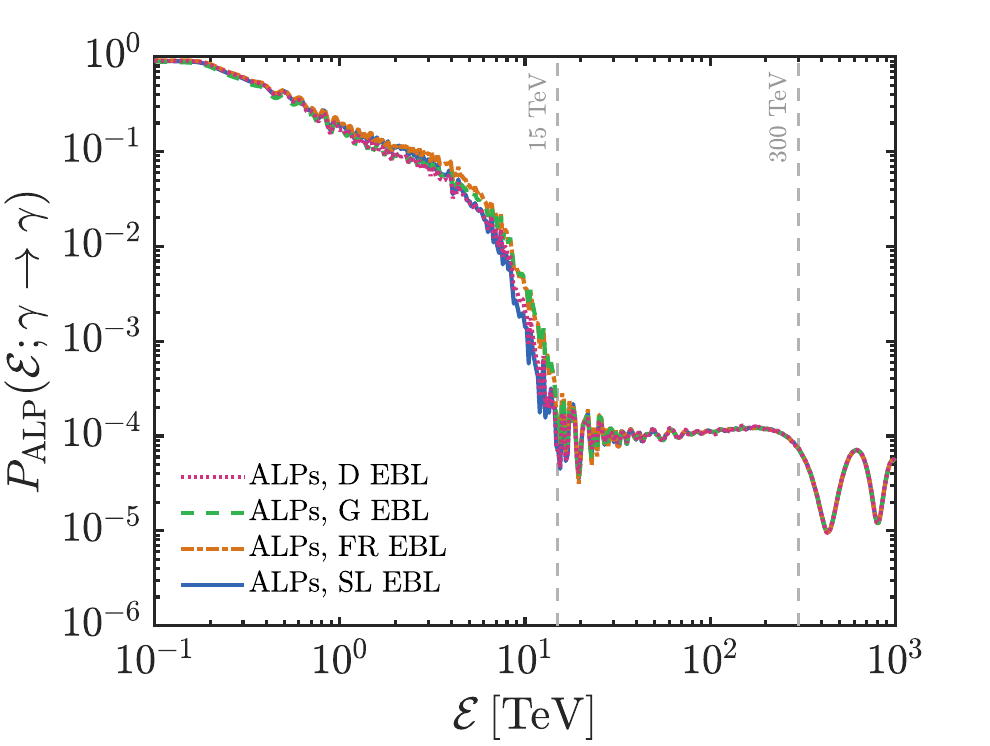}
\end{center}
\caption{\label{survProbSpiralALPsEBLlFIG} Photon survival probability $P_{\rm ALP} ({\cal E}; \gamma \to \gamma)$ (for $B_{\rm ext} = 10^{- 9} \, \rm G$) as a function of energy ${\cal E}$ in the case of a standard spiral host galaxy and adopting four different EBL models: 1) D EBL, 2) G EBL, 3) FR EBL, 4) SL EBL. We set the ALP parameters to $m_a=10^{-10} \, \rm eV$ and $g_{a\gamma\gamma}=4 \times 10^{-12} \, \rm GeV^{-1}$.}
\end{figure} 

\subsection{High host magnetic field}

\begin{figure*}[h]
\begin{center}
\includegraphics[width=.48\textwidth]{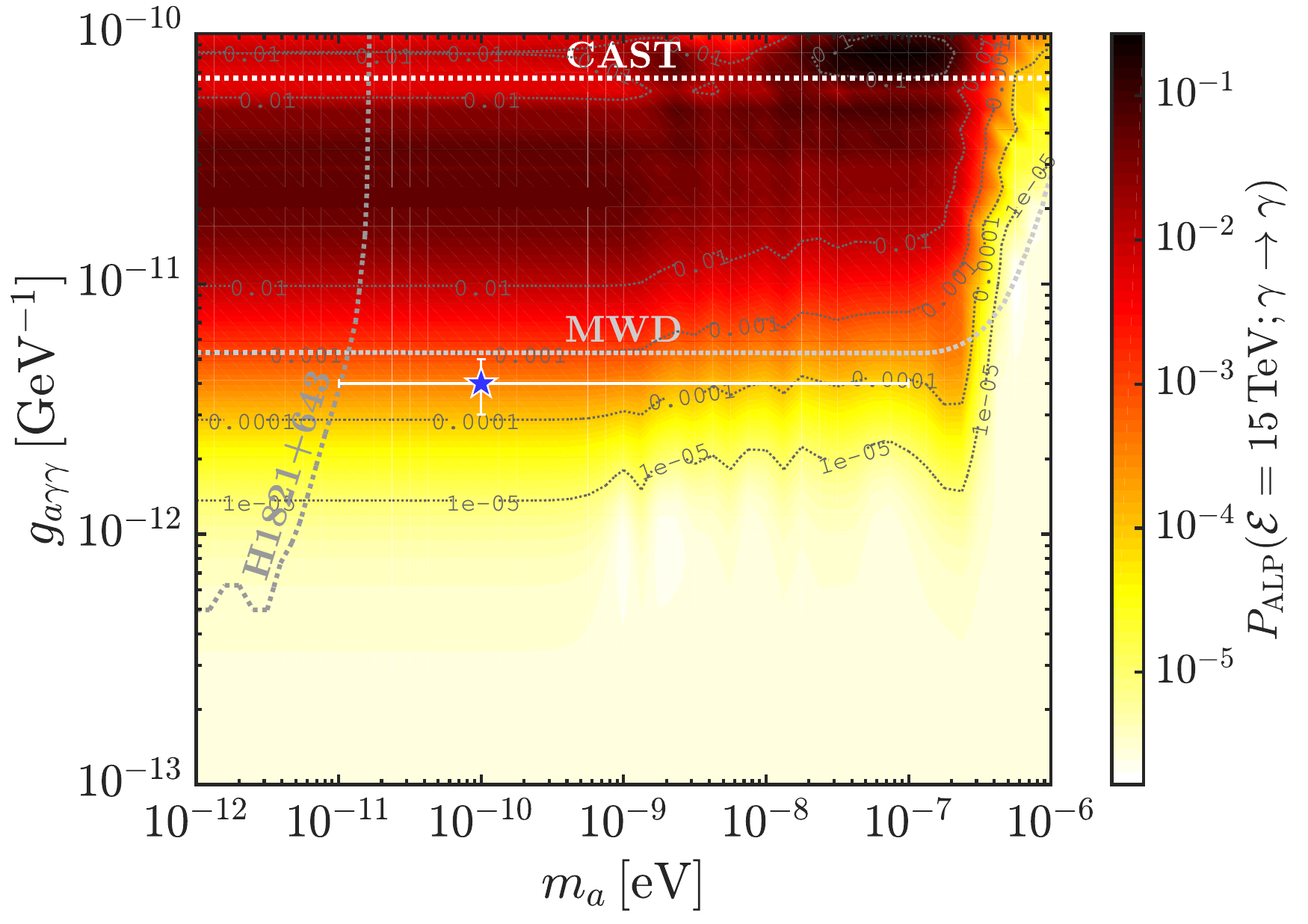} \hspace{3mm} \includegraphics[width=.48\textwidth]{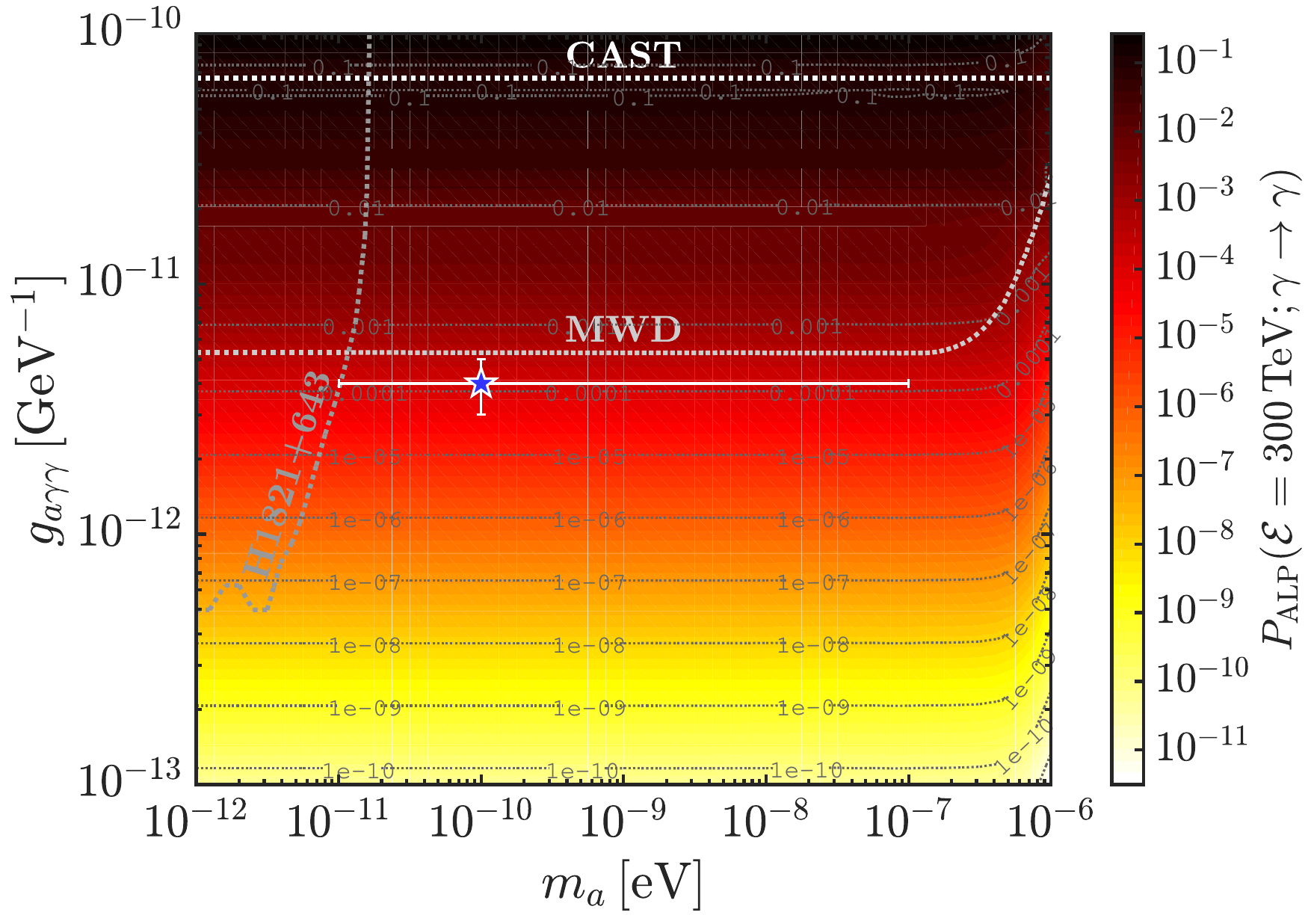}
\end{center}
\caption{\label{parSpaceMagnSpiralBig} Photon survival probability $P_{\rm ALP}({\cal E}; \gamma \to \gamma)$ for energies ${\cal E} = 15 \, \rm TeV$ (left panel) and ${\cal E} = 300 \, \rm TeV$ (right panel), as a function of the ALP mass $m_a$ and the photon-ALP coupling $g_{a\gamma\gamma}$. We assume the SL EBL model, $B_{\rm ext}=10^{- 9} \, \rm G$ and a highly magnetized spiral host galaxy. The blue star with error bar marks the choice for the ALP parameters: $m_a=10^{-10} \, \rm eV$ and $g_{a\gamma\gamma}=4 \times 10^{-12} \, \rm GeV^{-1}$. We also show the CAST bound~\cite{cast}, that arising from magnetic white dwarf (MWD) polarization~\cite{mwd} and the one obtained from H1821+643~\cite{limJulia}.}
\end{figure*}      

\begin{figure}[H]
\begin{center}
\includegraphics[width=.45\textwidth]{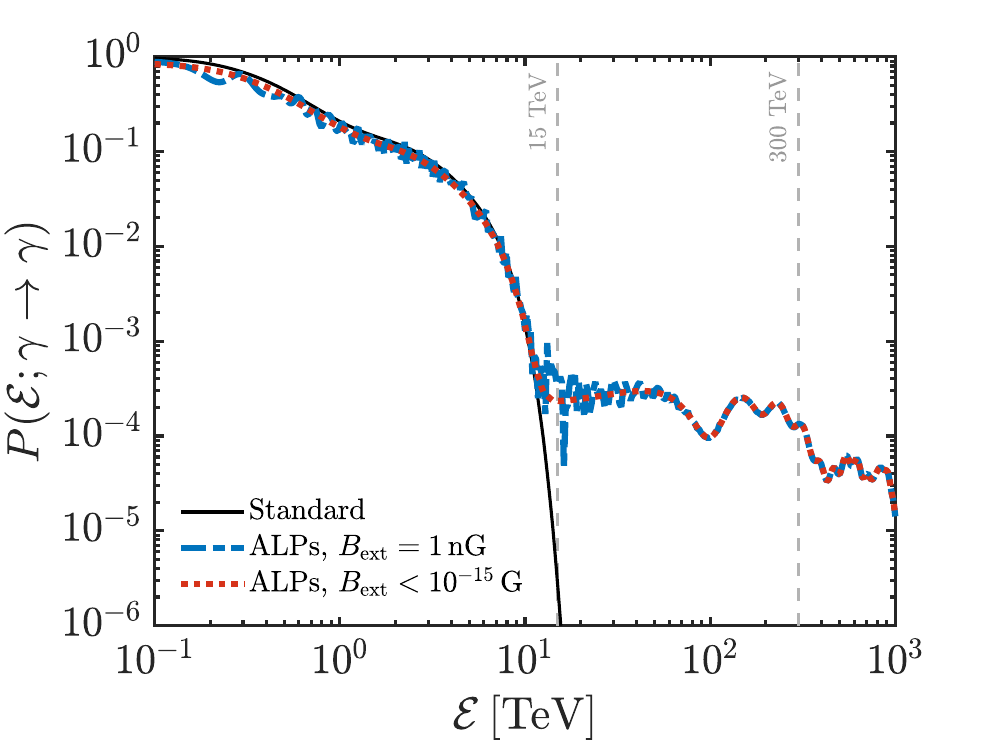}
\end{center}
\caption{\label{survProbFigMagnSpiral} Photon survival probability $P ({\cal E}; \gamma \to \gamma)$ as a function of energy ${\cal E}$ within conventional physics and including the photon-ALP oscillations (for $B_{\rm ext} = 10^{- 9} \, \rm G$ and $B_{\rm ext} < 10^{-15} \, \rm G$) in the case of a highly magnetized spiral host galaxy and adopting the SL EBL model. We set the ALP parameters to $m_a=10^{-10} \, \rm eV$ and $g_{a\gamma\gamma}=4 \times 10^{-12} \, \rm GeV^{-1}$.}
\end{figure} 

\begin{figure}[H]
\begin{center}
\includegraphics[width=.75\textwidth]{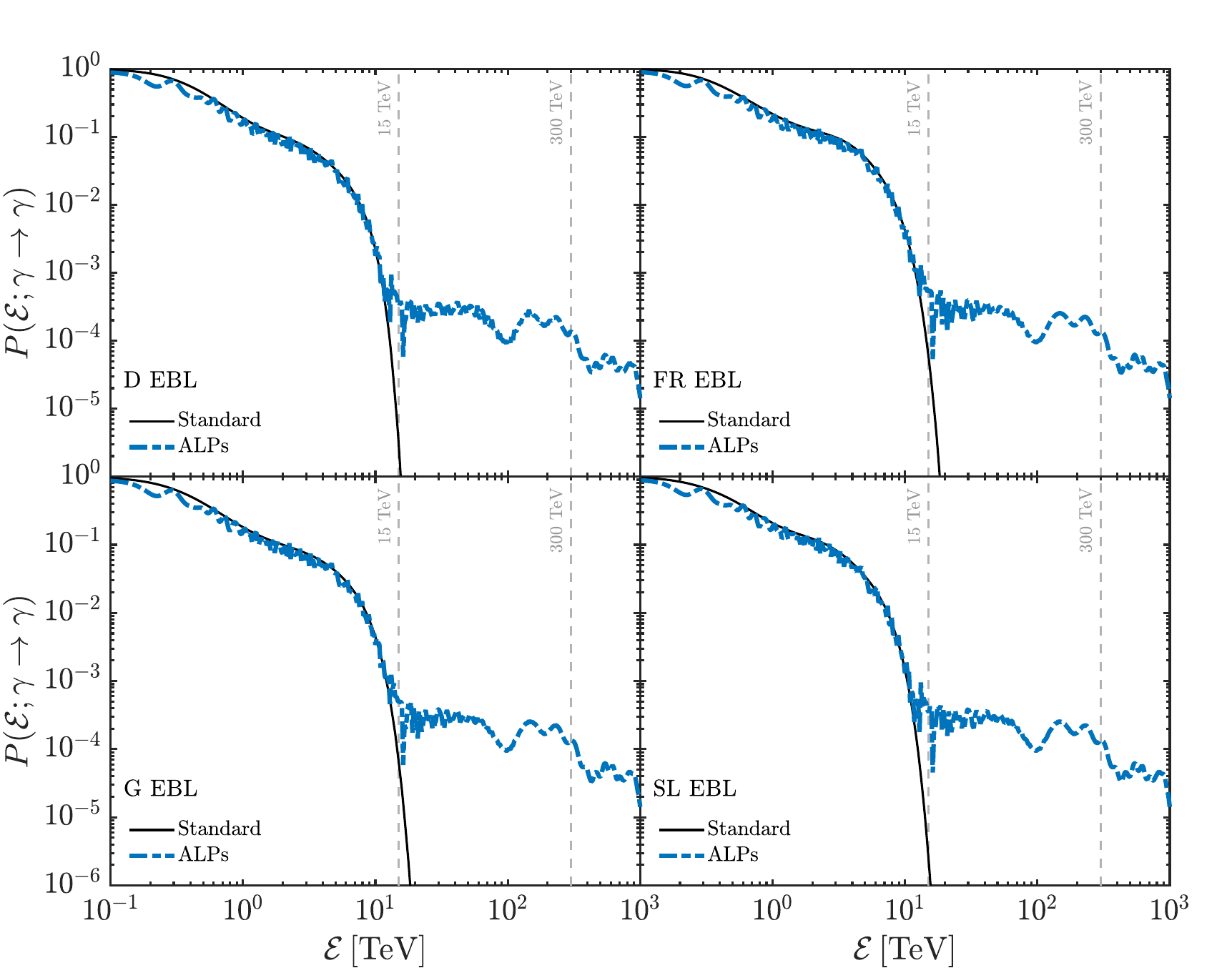}
\end{center}
\caption{\label{survProbMagnSpiralALPsEBLpanelFIG} Photon survival probability $P ({\cal E}; \gamma \to \gamma)$ as a function of energy ${\cal E}$ within conventional physics and including the photon-ALP oscillations (for $B_{\rm ext} = 10^{- 9} \, \rm G$) in the case of a highly magnetized spiral hosting galaxy and adopting four different EBL models: 1) D EBL (upper left panel), 2) G EBL (lower left panel), 3) FR EBL (upper right panel), 4) SL EBL (lower right panel). We set the ALP parameters to $m_a=10^{-10} \, \rm eV$ and $g_{a\gamma\gamma}=4 \times 10^{-12} \, \rm GeV^{-1}$.}
\end{figure} 

\begin{figure}[H]
\begin{center}
\includegraphics[width=.45\textwidth]{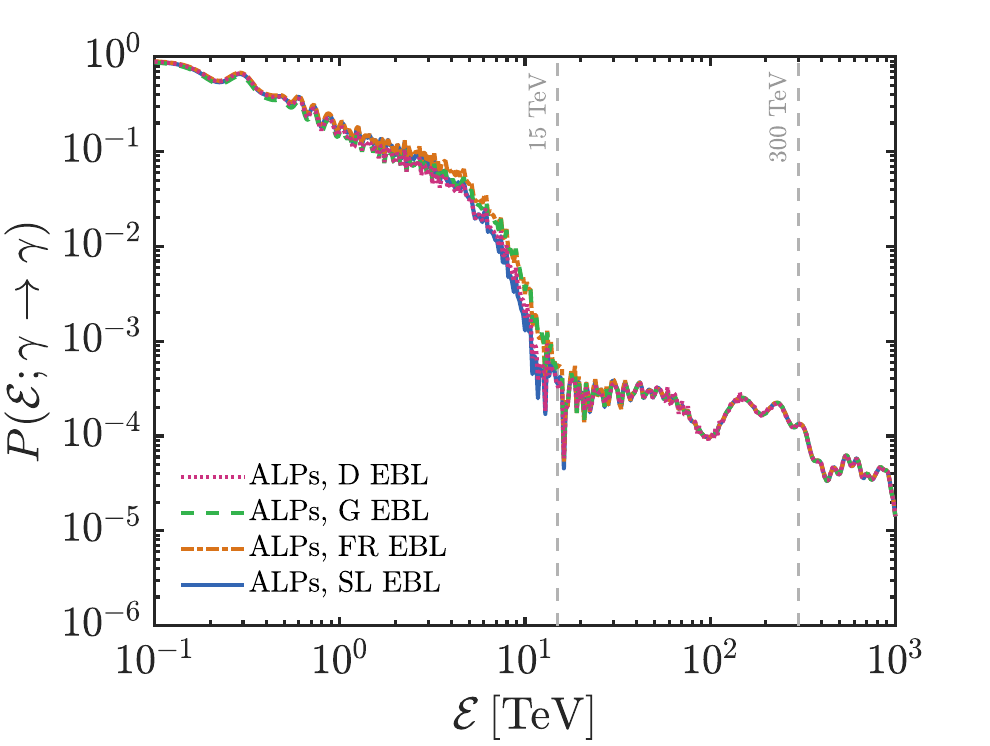}
\end{center}
\caption{\label{survProbMagnSpiralALPsEBLlFIG} Photon survival probability $P_{\rm ALP} ({\cal E}; \gamma \to \gamma)$ (for $B_{\rm ext} = 10^{- 9} \, \rm G$) as a function of energy ${\cal E}$ in the case of a highly magnetized spiral host galaxy and adopting four different EBL models: 1) D EBL, 2) G EBL, 3) FR EBL, 4) SL EBL. We set the ALP parameters to $m_a=10^{-10} \, \rm eV$ and $g_{a\gamma\gamma}=4 \times 10^{-12} \, \rm GeV^{-1}$.}
\end{figure} 

It should not come as a surprise that the curves pertaining to different EBL models are so similar: since in our case ALP effects outweigh those of EBL, the choice of its specific model becomes somewhat irrelevant (see Table I above). Having evaluated 
$P_{\rm ALP} ({\cal E}; \gamma \to \gamma)$ along with its involved uncertainties, we can apply Eq. (1) and we can compute the number of expected events $N_{\gamma}^{\rm ALP}$ observed by Carpet within the ALP scenario, indeed finding $N_{\gamma}^{\rm ALP} \sim 10^{- 5}$. 

\

\section{Previous constraints on LIV energy scales}

Throughout this Section we restrict our attention to the subluminal case (since the superluminal one is not considered in the main text). Hence photon decay and photon splitting do not occur. Here we consider the LIV bounds derived {\it before} the present work and concerning photons, which come from the observation of VHE photons emitted by astronomical sources at cosmological distances. We address them in a rather schematic fashion. It should be pointed out that in the literature the pair-production process is clouded by confusion since in some cases it means the Breit-Wheeler process whereas in other cases it is used for the Bethe-Heitler one. 

\

\noindent {\bf Time of flight} -- As explained in the main text, another implication of LIV theories is that the photon velocity becomes energy-dependent and in subluminal LIV photons are affected by a time delay. Because this effect increases with the covered distance, far-away powerful astronomical sources should be observed in order to discover this effect. So far only lower bounds on the LIV energy scales have been derived especially by means of the GRBs, which are the most suitable sources since the frequency of the emitted photons decreases rapidly soon after the onset of the afterglow~\cite{limitiLIV1,limitiLIV2,steckerglashowLIV,limitiLIV3,limitiLIV4,limitiLIV5,limitiLIV6,limitiLIV7,limitiLIV8,limitiLIV8a,limitiLIV9,limitiLIV9a,limitiLIV10,limitiLIV11,limitiLIV12,LIVlimLang,limitiLIV13,limitiLIV14,limitiLIV15,limitiLIV16}. The most stringent limits of this sort have been obtained from GRB 221009A and read ${\cal E}_{{\rm LIV}, \, 1} > 1.22 \times 10^{20} \, {\rm GeV}$ and ${\cal E}_{{\rm LIV}, \, 2} > 7.32 \times 10^{11} \, {\rm GeV}$ at $95 \%$ CL~\cite{limitiLIV17}.

\

\noindent {\bf Breit-Wheeler process} -- We recall that this is the pair-production process 
$\gamma + \gamma_{\rm EBL} \to e^+ + e^-$ with $\gamma_{\rm EBL}$ denoting an EBL photon. As discussed in Section III the uncertainty in the EBL level -- especially in the infrared -- affects any derived LIV bound. In addition, different constraints employ different EBL models, whose precision has changed over the years. These unquantified uncertainties reverberate on the inferred bounds.

As stated in the main text, subluminal LIV shifts the corresponding threshold upwards where the EBL level is lower, thereby making the Universe more transparent than expected to VHE photons. As first suggested in 2001 by Stecker and Glashow~\cite{steckerglashowLIV}, this effect can be used to put constraints on ${\cal E}_{{\rm LIV}, \, n}$ since the photon fluxes from extragalactic sources should be larger than predicted by conventional physics and so non-detection of enhanced fluxes constraints LIV (for Galactic sources this effect is irrelevant). Over the years this strategy has repeatedly been applied, and the improvement of the technological capabilities has made the resulting bounds progressively stronger. A fundamental step forward has been taken by Lang, Mart\'inez-Huerta and de Souza, who analyzed a data set of 111 energy spectra of 38 different sources measured by gamma-ray observatories~\cite{LIVlimLang}. Their limits at 95 $\%$ CL are ${\cal E}_{{\rm LIV}, \, 1} > 1.21 \times 10^{20} \, {\rm GeV}$ and ${\cal E}_{{\rm LIV}, \, 2} > 2.38 \times 10^{12} \, {\rm GeV}$ by assuming the EBL model of~\cite{frv2008}. These authors have also investigated the dependence of their results on the considered EBL model, finding ${\cal E}_{{\rm LIV}, \, 1} > 6.85 \times 10^{19} \, {\rm GeV}$ and ${\cal E}_{{\rm LIV}, \, 2} > 1.56 \times 10^{12} \, {\rm GeV}$ for the D EBL model~\cite{dominguez2011} and ${\cal E}_{{\rm LIV}, \, 1} > 1.50 \times 10^{20} \, {\rm GeV}$ and ${\cal E}_{{\rm LIV}, \, 2} > 2.17 \times 10^{12} \, {\rm GeV}$ for the G EBL model~\cite{gilmore2012}.

\

An important topic in this context is the Greisen–Zatsepin–Kuzmin (GZK) effect~\cite{GZK1,GZK2,GZK3}. Extragalactic ultra-high-energy (UHE) protons and heavier nuclei with energy up to ${\cal E} \simeq 10^{20} \, {\rm eV}$ scatter off the CMB producing neutral and charged pions. For the shake of our argument it is convenient to consider the simplest reaction $p + \gamma_{\rm CMB} \to p + \pi^0$ with the almost immediate production of UHE photons through the decay $\pi^0 \to \gamma_{\rm UHE} + 
\gamma_{\rm UHE}$. But no such UHE photons have  been detected on Earth, in spite of the fact that in subluminal LIV no EBL absorption occurs above a certain threshold. This fact has been used to put strong constraints on the LIV energy scale. While the idea is simple, its quantitative implementation is not, as it depends on many intrinsically undetermined variables, such as the composition of the incoming extragalactic charged UHE cosmic rays, their injection spectrum, the unknown extragalactic magnetic fields which adds up to the unknown distribution of their sources. As a result, the properties of the UHE charged cosmic rays which undergo the GZK process remains elusive in spite of the fact that several models have been proposed. The first LIV bound obtained by Galaverni and Sigl is ${\cal E}_{{\rm LIV}, \, 1} > 5.08 \times 10^{33} \, {\rm GeV}$ and ${\cal E}_{{\rm LIV}, \, 2} > 2.49 \times 10^{22} \, {\rm GeV}$~\cite{galaverni}. Subsequently, based on a more elaborated model for the UHE extragalactic cosmic rays Lang, Mart\'inez-Huerta and de Souza have derived the stronger LIV bound ${\cal E}_{{\rm LIV}, \, 1} > {\cal O}(10^{29}) \, {\rm GeV}$ and ${\cal E}_{{\rm LIV}, \, 2} > {\cal O}(10^{19}) \, {\rm GeV}$~\cite{lang2018}. But we shall see that this is not the end of the story. 

\

\noindent {\bf Bethe–Heitler process} -- This is a different pair-production process which 
-- at variance with the previous one -- is relevant only inside the Earth atmosphere~\cite{stanev}. When a very energetic photon interacts with the atmosphere it produces an $e^+ e^-$ pair in the Coulomb field of a nucleus $A$, namely according to the reaction $\gamma_{\rm UHE} + A \to e^+ + e^- + A$ whose cross section is~\cite{betheheitler} 
\begin{equation} 
\label{26012026a}
\sigma_{\rm BH} = \frac{28 Z^2 \alpha^3}{9 m_e^2} \left[{\rm ln} \left(\frac{183}{Z^{1/2}} \right) - \frac{1}{42} \right]~,
\end{equation}
where $\alpha$ is the fine structure constant and $Z$ is the nucleus charge. The first interaction of this kind is very important since it produces an electromagnetic shower. In the case of subluminal LIV the Bethe-Heitler cross section becomes~\cite{rss2017} 
\begin{equation} 
\label{26012026b}
\sigma_{{\rm BH, \, LIV}, n} \simeq \frac{16 Z^2 \alpha^3 {\cal E}_{{\rm LIV},n}^n}{3 {\cal E}^{n+2}} \, \ln \left(\frac{1}{\alpha Z^{1/3}} \right) \ln \left(\frac{{\cal E}^{n+2}}{2 m_e^2 {\cal E}_{{\rm LIV},n}^n} \right)~,
\end{equation} 
which is strongly suppressed with respect to $\sigma_{\rm BH}$. In order to find out its physical implications we consider the mean free paths $\lambda_{\gamma, \rm BH} (h)$ and $\lambda_{\gamma, {\rm BH, \, LIV}, n} (h)$ associated to $\sigma_{\rm BH}$ and 
$\sigma_{{\rm BH, \, LIV}, n}$, respectively, where $h$ is the hight of a generic nucleus of the atmosphere from the ground. Clearly we have 
\begin{equation}
\label{27012026a} 
\frac{\lambda_{\gamma, {\rm BH, \, LIV}, n} (h)}{\lambda_{\gamma, \rm BH} (h)} \simeq  \frac{\sigma_{\rm BH}}{\sigma_{{\rm BH, \, LIV}, n}}~.
\end{equation}
Since $\sigma_{{\rm BH, \, LIV}, n} \ll \sigma_{\rm BH}$ we get $\lambda_{\gamma, {\rm BH, \, LIV}, n} (h) \gg \lambda_{\gamma, \rm BH} (h)$, which means that the shower forms much deeper in the atmosphere. Therefore, when $\lambda_{\gamma, {\rm BH, \, LIV}} (h)$ exceeds a certain value depending on the experimental set-up the shower cannot be detected. More generally, a bound on the LIV energy scale arises by comparing the observed $\lambda_{\gamma, {\rm BH, \, LIV}, n} (h)$ to $\lambda_{\gamma, {\rm BH}} (h)$, which requires the knowledge of the emitted flux. This strategy has first been applied to the Crab nebula whose spectrum extends beyond $100 \, {\rm TeV}$ and is known very precisely, leading to ${\cal E}_{{\rm LIV}, \, 2} > 1.4 \times 10^{12} \, {\rm GeV}$ at $95 \%$ CL~\cite{satunin}. The subsequent application concerns the diffuse photon flux larger than $100 \, {\rm TeV}$ from the Galactic disk detected by the Tibet-AS$\gamma$ experiment~\cite{Tibet-AS} and the inferred bound is ${\cal E}_{{\rm LIV}, \, 2} > 1.7 \times 10^{13} \, {\rm GeV}$ at nominal $95 \%$ CL~\cite{satunin2}. However, the emitted flux is so poorly known that the real CL is much less than $95 \%$ and this bound {\it effectively disappears}.
 
\

Let us come back to the UHE photons produced by the GZK process within a LIV scenario. We have seen above that a strong LIV bound has been derived from the fact that they are not observed. We now inquire whether they can {\it actually} be observed, that is to say whether they can give rise to a photon-initiated air shower. To this end, we consider their LIV modified Bethe-Heitler mean free path 
\begin{equation}
\label{MFP}
\lambda_{\gamma, {\rm BH, \, LIV}, n} = \frac{1}{n_{\rm air} \, \sigma_{{\rm BH, \, LIV}, n}}~,
\end{equation}
where we take conservatively $n_{\rm air} = {\cal O}(10^{20}) \, \rm cm^{-3}$ the sea-level atmospheric number density and $Z = 7$  corresponding to nitrogen, which dominates the atmospheric composition. Assuming ${\cal E}_{{\rm LIV},1} = {\cal O}(10^{21}) \, \rm GeV$ and ${\cal E}_{{\rm LIV},2} = {\cal O}(10^{13}) \, \rm GeV$, and inserting Eq.~(\ref{26012026b}) into Eq.~(\ref{MFP}) we get $\lambda_{\gamma, {\rm BH, \, LIV}, 1} \ge {\cal O}(10^7) \, \rm km$ and $\lambda_{\gamma, {\rm BH, \, LIV}, 2} \ge {\cal O}(10^{10}) \, \rm km$, respectively. These results show that even if photons of energies 
${\cal E} \ge {\cal O}(10^{17}) \, \rm eV$ are produced through the GZK mechanism and survive propagation to the Earth due to LIV effects they are unable to interact with the atmosphere and so become invisible to air shower detection experiments. As a result, the previous LIV bounds based on the non-observation of UHE photons~\cite{galaverni,lang2018} {\it do not apply}. Still, for the energies considered in this Letter -- namely up to $\sim 300 \, \rm TeV$ -- photons in the 
${\rm ALP} + {\rm LIV}$ scenario have a mean free path ${\cal O}(1-10) \, \rm km$. So, they can instead interact with the atmosphere and initiate air showers which can be observed by experiments such as Carpet.

\

\section{ALP + LIV scenarios}

So far we have considered ALP and LIV effects separately. Because photon-ALP oscillations have been supposed to take place in a Lorentz-invariant framework, we have to investigate under which conditions they still work in the presence of LIV effects. 

As stated in the main text, the whole Lorentz-invariant setting is preserved by Lorentz invariance violation of the type considered here since it is assumed to conserve energy and momentum. Thus, the only change is that the dispersion relation of photons is modified by a LIV term, and in the main text we have seen that it looks like an effective mass square  
\begin{equation}
m^2_{\rm eff} \equiv - \, \xi_n \, \frac{{\cal E}^{n +2}}{{\cal E}^n_{{\rm LIV}, n}}~.
\label{29012026w}
\end{equation}
Because photon-ALP oscillations are energy-conserving, we assume that the ALP dispersion relation has just the LIV term (\ref{29012026w}) so that its looks almost like Eq. (5) of the main text, namely
\begin{equation}
p^2 = {\cal E}^2 \, \left(1 + \xi_n \, \frac{{\cal E}^n}{{\cal E}_{{\rm LIV}, n}^n} \right) - \, m^2_a~.
\label{29012026e}
\end{equation}
Otherwise stated, also the ALPs acquire an extra effective mass square equal to $m^2_{\rm eff}$ in Eq. (\ref{29012026w}). Altogether -- recalling Eq. (\ref{29012026q}) -- the same term 
\begin{equation}
\label{ALPandLIV1}
\Delta_{{\rm LIV}, n}({\cal E}) = \xi_n \, \frac{{\cal E}^{n+1}}{2 \, {\cal E}_{{\rm LIV}, n}^n}
\end{equation}
should be added to the three diagonal entries of ${\cal M} ({\cal E},y)$. As a consequence, the mixing matrix undergoes the transformation 
\begin{equation}
{\cal M} ({\cal E},y) \to {\cal M} ({\cal E},y) + \Delta_{{\rm LIV}, n}({\cal E}) \, I~,
\label{02022026q}
\end{equation}
where I denotes the $3 \times 3$ identity matrix. A look back at the beam propagation equation (\ref{propeq}) shows that Eq. (\ref{02022026q}) induces the following transformation of the wave function
\begin{equation}
\psi (y) \to {\rm exp} \bigl( i\, \Delta_{{\rm LIV}, n}({\cal E}) (y-y_0) \bigr) \, \psi (y)~. 
\label{02022026w}
\end{equation}
Manifestly Eq. (\ref{02022026w}) is a global phase change which leaves the physics of the photon-ALP sector unaffected. Thus, within the LIV models described above, where only kinematic effects are considered, photon-ALP oscillations remain unaffected. However, when considering the Breit-Wheeler process, the conventional mean free path $\lambda_{\gamma} ({\cal E}, y)$ in Eqs.~(\ref{deltaort}) and~(\ref{deltapar}) has to be replaced with its LIV-modified counterpart $\lambda_{\gamma, {\rm LIV}, n} ({\cal E}, y)$ for $n=1,2$, which can be computed from the LIV-induced optical depth $\tau_{\gamma, {\rm LIV},n} ({\cal E})$ as evaluated in~\cite{kifune1999,fairbairn2014,tavLIV2016} via the relation 
\begin{equation}
\lambda_{\gamma,{\rm LIV}, n} ({\cal E}, y) = \frac{y}{\tau_{\gamma, {\rm LIV},n} ({\cal E})}~.  
\label{28012026a}
\end{equation}
Thus, the only physical modification of ${\cal M} ({\cal E},y)$ arises from the replacement 
\begin{equation}
\lambda_{\gamma} ({\cal E}, y) \to \lambda_{\gamma,{\rm LIV}, n} ({\cal E}, y)~,  
\label{02022026a}
\end{equation}
and since ${\cal M} ({\cal E},y)$ retains exactly the same structure that it had in the absence of LIV we conclude that the ${\rm ALP} + {\rm LIV}$ scenario is physically consistent.

Finally, we emphasize that ALPs are affected by LIV, but LIV is unaffected by ALPs, magnetic fields or astrophysical effects.

\ 

\section{Other explanations for the observability of GRB 221009A}

After the proposal of the ${\rm ALP} + {\rm LIV}$ scenario to explain the observations of the VHE photons from GRB 221009A by LHAASO and Carpet~\cite{noi,grLIV}, a variety of models have been proposed to achieve the same goal. In the following, we provide a concise overview of the most representative models appeared so far. For clarity, the discussion will be schematic and brief.

\subsection{ALP models}

Given the strong interest in ALPs, numerous studies have explored their role in interpreting the LHAASO observations, generally following the approach first proposed in~\cite{noi}, albeit with variations in the ALP parameters, in the GRB 221009A properties, and in the computational techniques. However, there is a broad consensus that ALPs cannot account for the Carpet observation~\cite{bhm,CarenzaMarsh,troitskyGRB,wangboQalp,ZhangMa2023,troitsky2023,messicani}. For instance,~\cite{Nakagawa} introduces a scenario in which ALPs emerge in a non-perturbative setting associated with a first-order phase transition, while \cite{AvilaRojas} provides a detailed analysis of the GRB 221009A parameters within the ALP framework. But in both cases the constraints from the MWD bound~\cite{mwd} are not met. A similar drawback affects the model in~\cite{GaoBi}, which applies a novel probabilistic method to constrain all model parameters.

\subsection{Scalar model}

In contrast to ALPs, which are pseudoscalar in nature, a model involving a new singlet scalar particle has been proposed~\cite{Balaji}. This scalar interacts with Standard Model fields via a Higgs-like coupling, thereby implying that it could be abundantly produced in GRB 221009A. If the lifetime of the scalar allows it to decay into two photons sufficiently close to the Milky Way, they travel largely unaffected by the EBL thus providing a potential explanation of the LHAASO observations.

\subsection{Neutrino models}

In efforts to identify mechanisms that could reduce the EBL absorption of the highest-energy photons observed by LHAASO, two scenarios involving neutrinos have been suggested. 

\medskip

\noindent {\it First scenario} -- This postulates the existence of a heavy neutrino $N$ with a mass $m_N \sim 0.1 \, {\rm MeV}$, which can mix with pions and kaons emitted by GRB 221009A. Additionally, $N$ is assumed to couple to standard neutrinos while remaining decoupled from charged leptons. Further, the radiative decay $N \to \nu + \gamma$ is hypothesized to occur relatively near the Milky Way, hence allowing the resulting photons to reach us with very little EBL absorption. It is important to note that both the presence of $N$ and the radiative decay require new physics beyond the Standard Model~\cite{smirnovtrautner}. 

\medskip

\noindent {\it Second scenario} -- An alternative approach focuses first on the high-energy neutrinos generated in hadronic interactions, since GRB 221009A is expected to emit ultra-high-energy hadrons. During their propagation, these neutrinos may scatter off the cosmic neutrino background, thereby producing ALPs $a$ via the process $\nu + \nu \to a + a$. As a second step, the ALPs convert into photons in the Galactic magnetic field. A more intricate variant requires two distinct ALP species, $a$ and $a^{\prime}$, produced through $\nu + \nu \to a + a^{\prime}$. These models require not only the existence of ALPs but also the introduction of a dimension-five operator $a^2 \, \nu^T \, \nu$ within the Standard Model.~\cite{barnalfarzansmirkov}. 

\medskip

It should be emphasized that none of the proposed models so far address the detectability of the Carpet photon-like event. 

\subsection{Conventional physics models}

Two explanations based on standard physics alone have been proposed in order to account for the VHE photon emission associated with GRB 221009A.

\medskip

\noindent {\it {Proton beam scenario}} -- Independently of GRBs, a novel framework was introduced between 2010 and 2012 by Kusenko and collaborators as an alternative to the synchrotron-self Compton (SSH) mechanism of VHE emission from blazars. In this picture, a highly collimated beam of ultra-relativistic protons is accelerated within the jet and subsequently interacts with photons of the EBL, triggering electromagnetic cascades~\cite{EK2010,EKKBR2011,MDTM2012}. Since the secondary photons generated in this way are produced relatively close to the Earth, the effective EBL attenuation is strongly reduced. This mechanism requires the extragalactic magnetic field to be close to its lower observational bound considered in Section VI (Part C) in order to prevent significant beam broadening.

More recently, this framework has been applied to GRB 221009A with the aim of explaining the detection by LHAASO of VHE photons with ${\cal E} > 10 \, {\rm TeV}$ in terms of conventional physics, avoiding in this way more exotic possibilities such as ALPs~\cite{aronkus}. The authors also discuss a LIV interpretation, however -- as demonstrated in~\cite{noi} -- this option fails. A crucial assumption of the model is that the extragalactic magnetic field satisfies the stringent upper limit $B_{\rm ext} < 10^{-16} \, {\rm G}$. In addition to reproduce the observed photon spectrum, the model must ensure that the secondary photons arrive within the LHAASO observational time window of $2000 \, {\rm s}$, which applies to {\it all} detected photons regardless of energy.

Magnetic fields induce a time delay in the proton beam, which reverberates in the secondary photons. Three distinct contributions are relevant: (i) $\Delta t_{\rm host}$ due to the magnetic field ${\bf B}_{\rm host}$ of the host galaxy, (ii) $\Delta t_{\rm ext}$ associated with the extragalactic magnetic field ${\bf B}_{\rm ext}$, and (iii) $\Delta t_{\rm casc}$ arising from the magnetic field ${\bf B}_{\rm casc}$ within the electromagnetic cascade. The combined requirement reads
\[
\Delta t_{\rm host} + \Delta t_{\rm ext} + \Delta t_{\rm casc} < 2000 \, {\rm s}~.
\]

The host galaxy is assumed to be a spiral with a central regular magnetic field $B_{\rm host} \sim 1 \, {\mu}{\rm G}$ which decreases smoothly with the galactocentric distance and a correlation length $L_{\rm host} \sim 10 \, {\rm pc}$. By analogy with other GRB hosts~\cite{michalowski2015,deugartepostigo2024,thoene2024}, the location of GRB 221009A is assumed to be in a star-forming region in the outer spiral arms on the observer’s side of the galaxy, where the regular magnetic field is significantly weaker than $1 \, \mu{\rm G}$. The proton beam energy is taken as ${\cal E} \sim 10^5 \, {\rm TeV}$, and the distance traversed within the host is supposed to be $d_{\rm host} \sim 0.1 \, {\rm kpc}$, representative of the disk thickness. Under these assumptions, the host-induced time delay is~\cite{aronkus,dermerrazzaque2009}
\begin{equation}
\label{23102025a}
\Delta t_{\rm host} \sim 400 \, \left(\frac{d_{\rm host}}{{\rm kpc}} \, \frac{L_{\rm host}}{10 \, {\rm pc}} \right)^{3/2} \, \left(\frac{B_{\rm host}}{{\mu}{\rm G}} \, \frac{10^5 \, {\rm TeV}}{{\cal E}} \right)^2 \, {\rm s}~,
\end{equation}
which is comfortably below the $2000 \, \rm s$ threshold.

We argue that this conclusion is not justified. First, the comparison with the lower bounds on the extragalactic magnetic field in Section VI (Part C) shows that the adopted constraint 
$B_{\rm ext} < 10^{-16} \, {\rm G}$ is in tension with the most recent lower bound. More importantly -- as stated in Section I -- GRB 221009A is located at about $0.65 \, {\rm kpc}$ from the centre of a star-forming spiral galaxy viewed edge-on in first approximation~\cite{levan,blanchard} and {\it not} in its outskirts. In order to be very conservative, we take the host stellar radius ${\cal R}_{{\rm host}, *} \simeq 10 \, {\rm kpc}$ and in addition -- given the nearly edge-on orientation -- we suppose that the line of sight to GRB 221009A traverses the host disk for at least $d_{\rm host} \simeq 6 \, {\rm kpc}$, hence exceeding the value adopted in the model by a factor of $\sim 60$. Regarding the magnetic field, what matters is its behaviour within the disk ${\bf B}_{\rm disk}$. While typical spirals have $B_{\rm disk} \sim 10 \, {\mu}{\rm G}$ with correlation length $L_{\rm host} \sim 1 \, {\rm kpc}$~\cite{SpiralBrev,Heesen2023,vallee2011}, we adopt a conservative average value $\langle B_{{\rm host, disk}} \rangle \sim 0.1 \, {\mu}{\rm G}$ so that Eq.~(\ref{23102025a}) yields $\Delta t_{\rm host} \sim 6 \times 10^4 \, {\rm s}$, exceeding the LHAASO time window by roughly a factor of 30. Consequently, the proton beam scenario cannot account for the majority of the observed photons.

The discrepancy is in fact even more severe. As stressed in the main text, the observational challenge concerns photons with ${\cal E} > 10 \, {\rm TeV}$ -- as also emphasized by the authors of the present model -- while the relevant LHAASO time window for photons above ${\cal E} > 3 \, {\rm TeV}$ is only 670~$s$ (see main text). Hence, a viable model should require $\Delta t_{\rm host} < 670 \, {\rm s}$, whereas our conservative estimate yields $\Delta t_{\rm host} \sim 6 \times 10^4 \, {\rm s}$. We therefore conclude that a careful reassessment of the proton beam scenario as applied to GRB 221009A shows that it is doomed to failure.

\medskip

\noindent {\it {Neutron beam scenario}} -- This was originally proposed in~\cite{neutronbeam} and later revisited in~\cite{carpet3} as a potential explanation for the Carpet photon-like event. In this framework protons accelerated during the GRB prompt phase to energies ${\cal E} > 10^3 \, {\rm TeV}$ (in the observer frame) efficiently interact with the intense ambient radiation field, producing gamma rays, $e^+e^-$ pairs, neutrinos, and neutrons. The resulting neutrons next interact with the interstellar matter of the star-forming region surrounding the GRB, generating among other particles ultra-relativistic $e^+e^-$ pairs and neutrinos. The electrons and positrons then produce multi-TeV photons via synchrotron emission in the magnetic field of the host galaxy.

The maximum photon energy can be increased by approximately one order of magnitude either by assuming a host magnetic field an order of magnitude stronger than typically expected, or by invoking proton energies larger by a factor of $(3$--$4)$ than generally supposed. Under these conditions, photon energies up to $\sim 100 \, {\rm TeV}$ can be achieved~\cite{carpet3}. The delayed arrival time of the Carpet event is attributed to the angular dispersion of the neutron beam. Moreover, as discussed in~\cite{neutronbeam} the neutron--matter interactions inevitably produce a neutrino flux in the multi-TeV range, with an intensity comparable to that of the gamma-ray emission. However, IceCube upper limits~\cite{abbasiackermann2023,valtonenmattila2023} constrain this neutrino flux to be at least an order of magnitude smaller than the VHE fluence of GRB 221009A~{{(see also}~\cite{PiranLIV2}~{for further criticisms)}}. As a result, the neutron beam scenario appears unable to explain the Carpet event.

\end{widetext}


\begin{thebibliography}{9}
\bibitem{muons} Why muon detectors? They play a crucial role in discriminating between a primary photon emitted by the GRB 221009A and a secondary photon originating from proton-proton collisions in a cosmic ray shower, since in the latter case also pions are produced. We recall that neutral pions decay into two photons while charged pions decay into neutrinos and muons which can be recorded on Earth. So, the lack of muon detection by Carpet is a proof that their observed photon is not a misidentified hadron.
\bibitem{noz} Since we are dealing only with GRB 221009A the redshift-dependence of the various quantities will not be exhibited.
\bibitem{natun} Natural units ($c = 1$ and $\hbar = 1$) are employed throughout this Letter.
\bibitem{not} Observe that a LIV-deformed dispersion relation for ALPs is consistent with D-brane physics.
\end{thebibliography}
\end{document}